\documentclass[5p,times,twocolumn]{elsarticle}

\usepackage[a-1b]{pdfx}
\usepackage{xmpincl} 
\usepackage{hyperref}
\hypersetup{hidelinks}
\usepackage{amssymb}
\usepackage{booktabs,adjustbox} 
\usepackage[bracket-unit-denominator=false]{siunitx}
\usepackage{graphicx}
\usepackage[fleqn]{mathtools} 
\usepackage[UKenglish]{isodate}
\cleanlookdateon
\usepackage{caption,subcaption}
\usepackage{numcompress}

\emergencystretch=\maxdimen
\makeatletter
\def\ps@pprintTitle{%
   \let\@oddhead\@empty
   \let\@evenhead\@empty
   \let\@oddfoot\@empty
   \let\@evenfoot\@oddfoot
}
\makeatother

\begin{document}

\begin{frontmatter}
\title{Modelling ripple morphodynamics driven by colloidal deposition}
\author{James N. Hewett\corref{mycorrespondingauthor}}
\ead{james@hewett.nz}
\cortext[mycorrespondingauthor]{Corresponding author}
\author{Mathieu Sellier\corref{}}
\address{Department of Mechanical Engineering, University of Canterbury, Christchurch 8140, New Zealand}

\begin{abstract}
Fluid dynamics between a particle-laden flow and an evolving boundary are found in various contexts. We numerically simulated the morphodynamics of silica particle deposition from flowing water within geothermal heat exchangers using the arbitrary Lagrangian-Eulerian method. The silica particles were of colloidal size, with submicron diameters, which were primarily transported through the water via Brownian motion. First, we validated the Euler-Euler approach for modelling the transport and deposition of these colloidal particles within a fluid by comparing our simulation results with existing experiments of colloidal polystyrene deposition. Then we combined this multiphase model with a dynamic mesh model to track the gradually accumulated silica along the pipe walls of a heat exchanger. Surface roughness was modelled by prescribing sinusoidally-shaped protrusions on the wall boundary. The silica bed height grew quickest at the peaks of the ripples and the spacing between the protrusions remained relatively constant. The rough surface experienced a $\SI{20}{\percent}$ reduction in silica deposition when compared to a smooth surface. We also discuss the challenges of mesh deforming simulations with an emphasis on the mesh quality as the geometry changes over time.
\end{abstract}

\end{frontmatter}

\section{Introduction}
Deposition of particles onto a surface through a fluid is a common process found in both nature and industries. For example, sediment is transported by water in the ocean near the seabed and develops wave-formed sand ripples \citep{Hanes2001}. Deposited material may also detach from the seabed and the net amount of material changes over time, gradually modifying the seabed into ripple shaped structures \citep{Charru2013}. This phenomenon is also found in rivers \citep{Best2005} and channels \citep{Wynn2002}. Experimental \citep{Ouriemi2009} and numerical \citep{Arolla2015} studies have observed migration of these ripples downstream; by analysing sediment dynamics of particles with diameters on the order of $\SI{300}{\micro\metre}$. Ripple structures have also been observed in biofilms \citep{Purevdorj2002}, ski moguls \citep{Egger2002} and washboard roads \citep{Bitbol2009}. The coupling between the multiphase water and seabed forms an evolving boundary problem. A similar process of particle deposition and resuspension has been studied for industrial pipe flow \citep{Schaflinger1995,Stevenson2001}. An application of particle deposition with an evolving surface is 3-D printing \citep{Sachs1992} where manufacturing of more complex geometries can be achieved when compared to traditional tools.

The context of deposition with an evolving boundary we investigate in this paper is silica scaling which occurs in pipe flow within geothermal power stations. The silica is initially dissolved within water in the underground reservoirs and then, as the mixture reaches the plant equipment at the ground surface, the silica precipitates and forms silica particles \citep{Weres1981}. These colloidal (submicron diameter) silica particles are transported through pipe systems within the power plant and may attach to pipe surfaces; gradually accumulating a layer of silica deposit over the period of months or years \citep{Gunnarsson2005}. This layer of silica adversely affects the performance of heat exchangers. In particular: the pressure loss is increased due to the smaller pipe cross sectional area, and also the heat transfer is reduced as the layer of silica provides an additional thermal resistance.

Various silica scale morphologies have been observed in experiments \citep{Garibaldi1981,Brown2000,Dunstall2000} including fibrous deposits, cellular structures and rippled patterns. We numerically simulate the latter deposit structure where rippled structures are aligned normal to the direction of flow (circumferential direction for pipe flow). Numerical simulations and models on silica scaling have previously been explored \citep{Bohlmann1980,Pott1996,Moller1998,Masuri2012}. The focus of this study is the influence of the morphological evolution on the deposition rate by using an evolving boundary model.

One approach for modelling evolving boundary problems is the immersed boundary method \citep{Peskin1972} where the interface is modelled on a static mesh. An advantage of this approach is not requiring grid transformations \citep{Mittal2005}; these grid transformations, or remeshing, are computationally expensive tasks. However, a refined mesh is important for capturing momentum and concentration boundary layers and therefore a large proportion of the static grid would need to be refined in cases where the interface moves significantly; resulting in a high number of computational cells.

Another approach for modelling evolving boundary problems is tracking the boundary explicitly by modifying fluid properties of the finite volume cells. The block mesh method involves converting fluid cells to solid cells and the interface is defined as the boundary between the fluid and solid zones. This method has been used for modelling fouling in diesel engine exhaust systems \citep{Paz2013} and particle deposition on a cylinder in cross flow \citep{Hewett2015}. However, the boundary between the fluid and solid regions is restricted to the cell faces, resulting in a relatively coarse description of the interface. Furthermore, the resolution of the boundary layer is also restricted by the refinement of the mesh and an excessive number of cells would be required for significant deformations; a similar disadvantage to the immersed boundary method.

A third evolving boundary model is the arbitrary Lagrangian-Eulerian (ALE) method \citep{Hirt1974,Hughes1981,Donea2004} which transforms the mesh throughout the simulation to facilitate the boundary deformation. The advantage of this method is that the mesh topology remains constant throughout the simulation and only the individual cells are transformed. The majority of the deformation can be absorbed in the far field region allowing the mesh near the boundaries to remain mostly consistent for the duration of the simulation, and consequently preserving the resolution of the flow gradients in the boundary layer. Fluent, a computational fluid dynamics software, has developed a dynamic mesh model where flows are simulated with a dynamic domain by using deforming boundaries. The boundaries are prescribed with user-defined functions and the mesh interior is dynamically updated at each time step. This model has been validated against experiments with a heart valve \citep{Dumont2004}, with the erosion of a cylinder in cross flow \citep{Hewett2017} and the melting ice front around a heated cylinder \citep{Hewett2017a}. In this paper, we use the dynamic mesh model, coupled with modelling the silica particle phase, to explore the impact of boundary evolution on accumulation of colloidal silica in pipe flow at the microscale and compare our results with an experimental test rig \citep{Sinclair2012,Kokhanenko2015}.

\section{Methods}
Two flow configurations were investigated: Poiseuille flow (parallel plates with stationary walls) and Couette flow (parallel plates with the upper wall moving and the lower wall stationary). Both configurations had distinct geometry and particle parameters but had similar physicochemical properties and dimensionless flow dynamics including steady 2-D laminar flow conditions. First, the Poiseuille flow case was used as a validation using existing experimental data to ensure the chosen method of simulating colloidal particle transport in laminar flow was accurate. Polystyrene colloidal particles were modelled within a parallel plate flow cell and their initial deposition rate onto the smooth rigid boundary was compared with experimental \citep{Yang1999} and analytical solutions \citep{Elimelech1995}. Second, the accumulation of deposited colloidal silica particles was modelled in turbulent pipe flow where the pipe wall augmented as a function of the silica deposition flux; causing an evolving boundary problem. The pipe wall had a combination of smooth and sinusoidally-shaped protrusions: a single protrusion was simulated and then a series of bumps, forming a surface roughness model. Both configurations were simulated using two phases: one for the fluid, and a second for the dilute suspension of particles. ANSYS Fluent R17.0 was used as the finite volume solver and MATLAB R2016b for the data analysis and visualisation.

\subsection{Geometry and meshing}
Dimensionless length scales such as the dimensionless vertical wall distance $y^+$ are defined as
\begin{equation}
y^+ = \dfrac{y u^*}{\nu}
\end{equation}
where $y$ is the vertical distance normal to the surface, $u^*$ the shear velocity and $\nu$ the kinematic viscosity. Time $t$ is non-dimensionalised with%
\begin{equation}
t^+ = \frac{t {u^*}^2}{\nu}
\end{equation}
and the shear velocity is defined as
\begin{equation}
u^* = \sqrt{\frac{\tau_{w,\text{smooth}}}{\rho}}
\end{equation}
where $\tau_{w,\text{smooth}}$ is the wall shear stress in the absence of any sinusoidal protrusions and $\rho$ the fluid density.

\subsubsection{Poiseuille flow}
The experimental test section \citep{Yang1999} had a length of $L = \SI{76}{\milli\metre}$. Half the distance between the parallel plates was $b = \SI{0.3}{\milli\metre}$, and we set an arbitrary depth of $\SI{2}{\milli\metre}$ using one cell in the spanwise direction (across the plates, normal to the flow). We extended the length of the computational domain by $\SI{20}{\milli\metre}$ ($\approx \num{67}b$) in both the upstream and downstream directions in our computational domain to isolate end effects of the inlet and outlet. These developing flow regions were excluded from data analysis and are truncated from the presented results for clarity. 

A structured Cartesian grid was generated with uniform cell spacings in the streamwise direction and a bias near the walls in the wall normal direction (between the parallel plates) to capture the near-wall momentum and concentration gradients. An expansion ratio of five was used in the wall normal direction such that the edge lengths of the centre cells were five times that of the near-wall cell widths. Mesh resolutions are listed in Table~\ref{tab:TableA} alongside the results of the mesh convergence analysis.

\begin{table}[t]
\centering\small
\caption{Mesh convergence ($\text{streamwise} \times \text{wall normal}$) and validation of particle deposition flux at the centre of the plate, for the Poiseuille flow configuration, compared with the analytical Smoluchowski-Levich approximation.}
\label{tab:TableA}
\begin{tabular}{@{}lS[table-format=3.1]S[table-format=2.1]@{}}
\toprule
Mesh                & {$j~(\SI{}{\#\per\centi\metre\squared~\second})$} & {Error (\SI{}{\percent})} \\ \midrule
$116 \times 10$     & 249.0        & -17.3     \\
$145 \times 30$     & 312.1        & 3.7       \\
$232 \times 50$     & 301.9        & 0.3       \\
$464 \times 70$     & 300.4        & -0.2      \\
Analytic            & 301.1        &           \\ \bottomrule
\end{tabular}
\end{table}

\subsubsection{Couette flow}\label{sec:SectionA}
The experimental test rig \citep{Sinclair2012,Kokhanenko2015} recirculated fluid containing a dilute suspension of silica nanoparticles through a system of pipes. Silica deposition was observed along a straight pipe section of length $L = \SI{1}{\metre}$ and diameter $D = \SI{15}{\milli\metre}$. The spatial size of the silica ripples had a height of $h < \SI{50}{\micro\metre}$ which is significantly smaller than that of the pipe diameter. The region of interest for this paper is the deposited silica scale which gradually accumulates over time. Therefore a computational subdomain surrounding a ripple was established to avoid unnecessary computations of the far field flow. We assumed that the effect of the evolving ripple had a negligible impact on the bulk flow in the pipe: the blockage ratio is on the order of less than $\SI{0.7}{\percent}$.

The subdomain had a length of $L_\text{sub}^+ = \num{50}$ and height $H_\text{sub}^+ = \num{10}$. The length was chosen such that the flow upstream and downstream of the ripple developed prior to entering or exiting the domain respectively. The flow simulations around the micron sized ripples occur in the laminar viscous sublayer of the turbulent pipe flow and therefore the approximation of modelling this case as Couette flow is valid for up to $y^+ \approx 10$. Further details on assumptions and reasoning are provided in Section~\ref{sec:SectionB}.

\begin{figure}[t] \centering
  \includegraphics[scale=0.75]{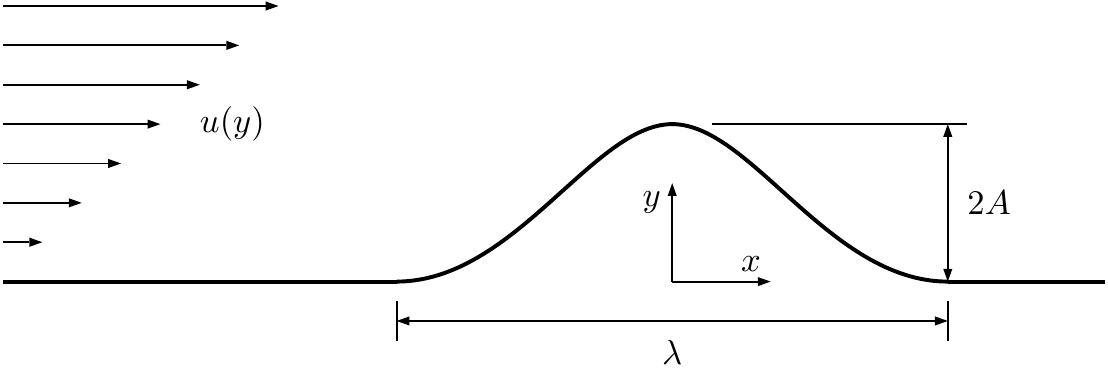}
  \caption{Schematic of a single ripple showing the height $2A$ and length $\lambda$, along with a description of the coordinate system. The particle-laden fluid is flowing from left to right across the ripple, and the velocity profile shown is applied far upstream of the ripple.\label{fig:Figure1}}
\end{figure}

\begin{figure}[t] \centering
  \begin{subfigure}{\linewidth}\begin{flushright}
    \includegraphics{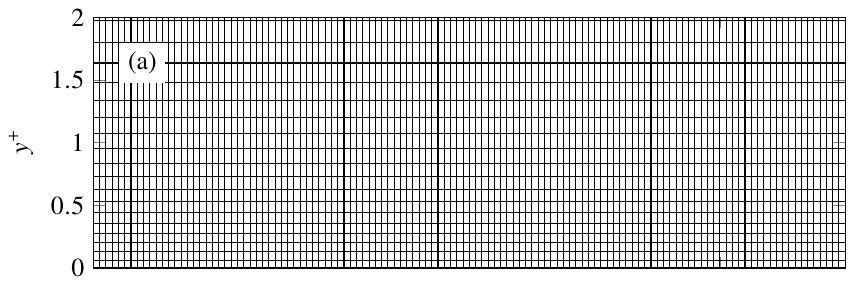}\hspace*{1.85pt}%
    \phantomcaption{\label{fig:Figure2a}}
  \end{flushright}\end{subfigure}\hfill%
  \begin{subfigure}{\linewidth}\begin{flushright}
    \includegraphics{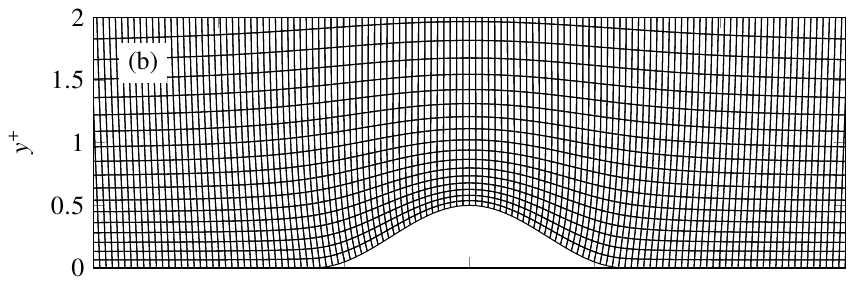}\hspace*{1.85pt}%
    \phantomcaption{\label{fig:Figure2b}}
  \end{flushright}\end{subfigure}\hfill%
  \begin{subfigure}{\linewidth}\begin{flushright}
    \includegraphics{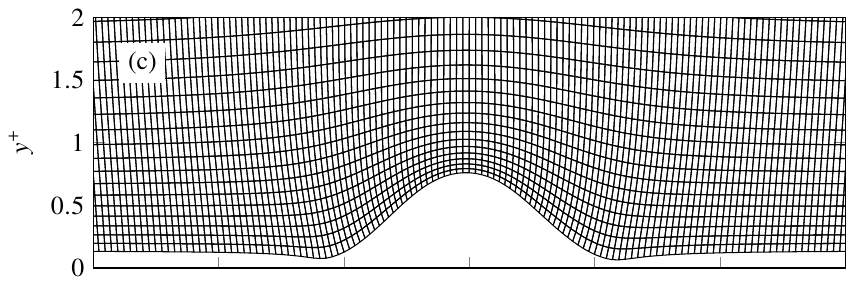}\hspace*{1.85pt}%
    \phantomcaption{\label{fig:Figure2c}}
  \end{flushright}\end{subfigure}\hfill%
  \begin{subfigure}{\linewidth}\begin{flushright}
    \includegraphics{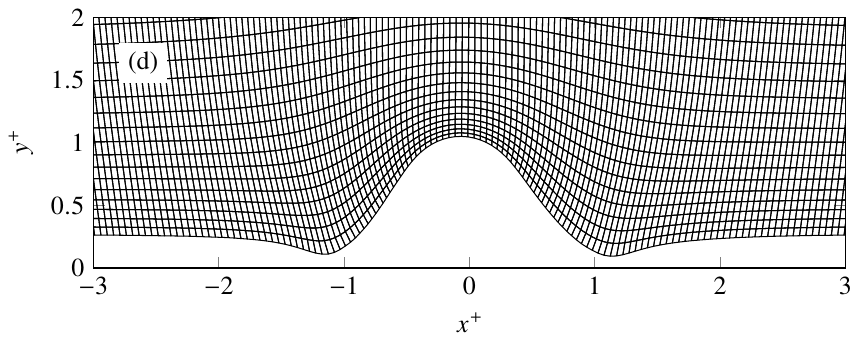}%
    \phantomcaption{\label{fig:Figure2d}}
  \end{flushright}\end{subfigure}
  \caption{Close-up snapshots of the computational mesh for the simulation of a single evolving bump in the laminar viscous sublayer at: (\subref{fig:Figure2a}) the initially generated mesh; (\subref{fig:Figure2b}) $t^+ = 0$; (\subref{fig:Figure2c}) $t^+ = 5 \Delta t^+$; and (\subref{fig:Figure2d}) $t^+ = 10 \Delta t^+$ where $\Delta t^+ = \num{2500}$.\label{fig:Figure2}}
\end{figure}

The initial profile of the single bump on the lower boundary was applied as a sinusoidal shape with
\begin{equation}\label{eqn:EquationL}
y^+ = A + A \cos (k x^+) \text{ for } -\frac{\lambda}{2} < x^+ < \frac{\lambda}{2}
\end{equation}
where $A = 0.25$ is the amplitude and $k = 2 \pi / \lambda$ the wavenumber with $\lambda = 2.5$ the wavelength of the sinusoid, as shown in Figure~\ref{fig:Figure1}. The bump had an initial maximum height of $h_0^+ = 2A$ and length $l_0^+ = \lambda$ as shown in Figure~\ref{fig:Figure2b}; giving an aspect ratio of $l_0^+/h_0^+ = \num{5}$. For the surface roughness model with five bumps, this equation was also applied but with the streamwise limits of $x^+ \pm 5 \lambda / 2$.

Similar to the Poiseuille flow case, the Couette flow mesh was a structured Cartesian grid with uniform cell spacings in the streamwise direction, but only a bias toward the lower boundary in the wall normal direction (with an expansion ratio of ten). Mesh convergence for the Couette flow case was achieved as shown in Figure~\ref{fig:Figure5} and a close-up of the mesh near the lower boundary is shown in Figure~\ref{fig:Figure2a}.

\subsection{Fluid description}
The Navier-Stokes equations with the ALE formulation \citep{Hirt1974,Donea2004} were used to model the fluid motion in the simulations and included the effect from the evolving boundary. The fluid was assumed Newtonian and laminar. The ALE momentum equation for an arbitrary control volume $V$ with surface $S$ with these assumptions is
\begin{equation}
\begin{split}
&\frac{d}{dt} \int_V \rho \textbf{u} ~dV + \int_S \rho \textbf{u} (\textbf{u} - \textbf{u}_S) \cdot \hat{\textbf{n}} ~dS = - \int_V \nabla p ~dV \\&\qquad + \int_S \mu (\nabla \textbf{u} + (\nabla \textbf{u})^T) \cdot \hat{\textbf{n}} ~dS
\end{split}
\end{equation}
where $\textbf{u}$ is the fluid velocity, $\textbf{u}_S$ the surface velocity of the control volume, $p$ the pressure, $\mu$ the dynamic viscosity and $\hat{\textbf{n}}$ the unit normal vector pointed outward from the boundary. Similarly, the continuity equation is
\begin{equation}
\frac{d}{dt} \int_V \rho ~dV + \int_S \rho (\textbf{u} - \textbf{u}_S) \cdot \hat{\textbf{n}} ~dS = 0
\end{equation}

The ALE formulation of the Navier-Stokes equations were employed in the simulations using the dynamic mesh model of Fluent to ensure that the effect on the fluid due to the motion of the moving boundary was included. Although, because the ratio of the depositing silica velocity with the fluid velocity was approximately $\mathcal{O}(10^{-6})$, the motion of the accumulating silica deposit on the pipe wall had a negligible impact on the flow. Furthermore, the flow field was practically in steady state because the flow was steady for each mesh step, and the evolving fluid-silica interface moved at a significantly slower rate compared to the pipe flow. Lastly, the fluid was assumed to be incompressible. Therefore, the governing equations can effectively be expressed in their Eulerian forms for each mesh step, with the momentum equation
\begin{equation}
\rho (\textbf{u} \cdot \nabla \textbf{u}) = - \nabla p + \mu \nabla \cdot (\nabla \textbf{u} + (\nabla \textbf{u})^\text{T})
\end{equation}
and the continuity equation
\begin{equation}
\nabla \cdot \textbf{u} = 0
\end{equation}

A cell-centred finite volume method was employed for solving the transport equations of the fluid phase. Pressure and velocity were coupled with the algorithm SIMPLE: semi-implicit method for pressure-linked equations \citep{Patankar1972}, involving an iterative process. The spatial discretisation scheme used for the gradients was least squares cell based, pressure was second-order upwind and momentum third-order MUSCL: monotonic upstream-centred scheme for conservation laws \citep{vanLeer1979}. Under-relaxation factors for pressure and momentum were used with values of $\num{0.3}$ and $\num{0.7}$ respectively. All flow field values were initialised with the inlet boundary condition.

The geometry changed over time in the Couette flow cases as the fluid-silica interface evolved. The flow field was treated as steady state for each discrete step of the mesh deformation process. However, evolution of the interface occurred over time and thus the model was in a pseudo steady state. The transient solver in Fluent was employed such that each simulated time step corresponded to a discrete mesh step; allowing the complete interface evolution to be run as one simulation. Steady state for each discrete mesh update was achieved by setting an arbitrarily large time step. This approach was verified by comparing results from the steady state solver and the transient solver with this very large time step.

\subsubsection{Poiseuille flow boundary conditions}
No slip conditions were applied to the upper and lower boundaries. Symmetry was applied on the front and back boundaries: forcing the 2-D assumption. A uniform inlet velocity $u_\text{inlet}$ was applied on the upstream boundary (the Poiseuille velocity profile was fully developed after the entrance region) and a pressure outlet on the downstream boundary.

\subsubsection{Couette flow boundary conditions}\label{sec:SectionB}
A no slip condition was also applied on the lower boundary in the pipe flow configuration: the fluid-silica interface. Similar to the Poiseuille case, a symmetry condition was imposed on the front and back boundaries, and a pressure outlet was applied on the downstream boundary. A velocity inlet condition was used for the upstream boundary with a linear velocity profile following
\begin{equation}
\textbf{u} = u^* y^+ {\hat{\textbf{\i}}}
\end{equation}
where ${\hat{\textbf{\i}}}$ is the unit normal vector in the streamwise direction.

The top boundary, approximately representing the edge of the laminar viscous sublayer, had a no slip condition applied with a moving wall motion of this velocity.

The viscous sublayer rises as the silica accumulates and shifts the interface as shown in Figure~\ref{fig:Figure2c}. Consequently, the $y^+$ value was offset by the far field bed height. This tracking of the sublayer as the bed rises assumes that this sublayer remains consistent throughout the simulation and has a negligible effect on the bulk pipe flow; the ratio of this sublayer and the pipe diameter was $\mathcal{O}(10^{-4})$.

\subsection{Particle phase}
Particles were tracked in the Eulerian reference frame as a dilute mixture (no particle-particle interactions) with negligible inertial effects (colloidal size) and in steady state. Fick's law of diffusion is applicable to the colloidal particles of interest in this paper, and a common transport equation \citep{Guha2008} for these conditions is the convection-diffusion equation
\begin{equation}\label{eqn:EquationO}
\nabla \cdot (c^+ \textbf{u} - D_B \nabla c^+) = 0
\end{equation}
where $D_B$ is the Brownian diffusion, $c^+ = c/c_\text{bulk}$ the dimensionless particle concentration with $c$ the particle concentration and $c_\text{bulk}$ the bulk particle concentration. Brownian diffusion is defined as
\begin{equation}\label{eqn:EquationC}
D_B = \dfrac{k_B T}{3 \pi \mu d_p}
\end{equation}
where $k_B = \SI{1.38e-23}{\joule\per\kelvin}$ is the Boltzmann constant, $T = \SI{295.15}{\kelvin}$ the temperature and $d_p$ the particle diameter.

The transport equation for the particle phase (Equation~\ref{eqn:EquationO}) was also discretised with the finite volume method using a third-order MUSCL algorithm, and the initial dimensionless concentration of the field was $c_0^+ = 0$.

Particle relaxation time $\tau_p$ is a measure of their response to the surrounding fluid and is defined in the Stokes regime as
\begin{equation}
\tau_p = \frac{\rho_p {d_p}^2}{18 \mu}
\end{equation}
where $\rho_p$ is the particle density. The mechanisms responsible for particle transport are typically characterised by using a dimensionless form of this particle relaxation time, $\tau_p^+$, non-dimensionalised like $t^+$.

The dimensionless rate of deposition of particles on a surface is defined as
\begin{equation}
u_d^+ = \frac{j}{c_\text{bulk} u^*}
\end{equation}
where $j$ is the particle flux at the boundary and is evaluated with
\begin{equation}\label{eqn:EquationA}
j = D_B \frac{dc}{dn}
\end{equation}
where the concentration gradient is taken in the normal direction to the lower plate (Poiseuille flow) or local fluid-silica interface (Couette flow) boundary.

\subsubsection{Poiseuille flow particle boundary conditions}
Boundary values for the particle phase were $c^+ = 0$ on the two parallel plates (perfect-sink model \citep{Elimelech1995}) and $c^+ = 1$ at the inlet. The entrance region of the domain upstream of the two plates, where the flow field develops, also had $c^+ = 1$ imposed such that the dimensionless particle concentration at the start of the plates was set at unity. The outlet boundary had a zero diffusive particle flux applied ($dc^+/dn^+ = 0$).

\subsubsection{Couette flow particle boundary conditions}
Similar to the Poiseuille flow, the lower boundary (fluid-silica interface) was $c^+ = 0$, whereas the upper boundary (edge of the viscous sublayer) had $c^+ = 1$ as the bulk of the pipe flow was approximately uniform. The outlet boundary had a zero diffusive particle flux applied ($dc^+/dn^+ = 0$).

The inlet particle concentration profile was obtained by first simulating the full test pipe length of $L = \SI{1}{\metre}$ with an inlet concentration of $c^+ = 1$. A slice of the concentration profile at $x = \SI{0.5}{\metre}$ was then extracted and applied to the short subdomain; linearly interpolating from the profile points to the subdomain mesh.

\subsection{Deforming mesh}\label{sec:SectionC}
The mesh was deformed in the Couette flow cases. Gradual accumulation of material on the wall surfaces were due to the silica particle deposition. The multiphase mixture of water and silica particles were in kinetic equilibrium such that the chemical reaction rates at the interface were quick compared to the flow time; growth of the silica deposit was mass transfer controlled.

The deposition rate of these silica colloids can be related to the particle concentration in the Eulerian reference frame by calculating the concentration gradient at the boundary. The interface velocity of the accumulating boundary is given by Fick's law for the solute concentration field, analogous to the case of dissolution \citep{Moore2017}, as
\begin{equation}\label{eqn:EquationD}
\textbf{v}_n = D_B \dfrac{d c^+}{dn} \hat{\textbf{n}}
\end{equation}
where $\hat{\textbf{n}}$ is the unit normal pointed towards the fluid domain.

The computational nodes on the lower boundary (deforming fluid-silica interface) were displaced with
\begin{equation}\label{eqn:EquationF}
\Delta \textbf{x} = \Delta t \textbf{v}_n
\end{equation}
where $\Delta t$ is the time step between mesh updates.

The remaining boundaries (on the inlet, outlet and upper sides) were marched upwards such that the computational domain tracked with the evolving silica bed. These nodes were marched by offsetting their positions by the displacement of the left-most interface boundary node (representative of the far field silica bed height).

The motion of the interior nodes were governed by a linearly elastic solid model. Mesh motion was governed by
\begin{equation}\label{eqn:EquationGa}
\nabla \cdot \underline{\boldsymbol{\sigma}} (\textbf{m}) = \underline{\textbf{0}}
\end{equation}
where $\underline{\boldsymbol{\sigma}}$ is the stress tensor and $\textbf{m}$ the mesh displacement vector,
\begin{equation}\label{eqn:EquationGb}
\underline{\boldsymbol{\sigma}}(\textbf{m}) = E \text{tr}( \underline{\boldsymbol{\epsilon}}(\textbf{m})) \underline{\textbf{I}} + 2 G \underline{\boldsymbol{\epsilon}} (\textbf{m})
\end{equation}
where $\underline{\boldsymbol{\epsilon}}$ is the strain tensor, $G$ the shear modulus and $E$ the Young's modulus,
\begin{equation}\label{eqn:EquationGc}
\underline{\boldsymbol{\epsilon}} (\textbf{m}) = \frac{1}{2} (\nabla \textbf{m} + (\nabla \textbf{m})^\text{T})
\end{equation}

The solution of Equations~\ref{eqn:EquationGa} to \ref{eqn:EquationGc} depends only on the ratio of $G$ and $E$ and this ratio was parameterised with Poisson's ratio
\begin{equation}
\text{Po} = \frac{1}{2(1+G/E)}
\end{equation}

These mesh deformation equations were employed by enabling the smoothing method of the dynamic mesh model in Fluent. Deposition, and consequently the deforming boundary, is a function of the concentration gradient (Equation~\ref{eqn:EquationD}) and therefore the concentration boundary layer must be converged throughout the evolving simulation. A Poisson's ratio of $\text{Po} = \num{0}$ was chosen as this value provided the best mesh evolution considering that a constant resolution of the concentration gradient in the boundary layer was ideal.

\subsection{Node shuffle algorithm}\label{sec:SectionD}
Finite volume cells were both modified and translated throughout the simulations as the moving boundary deformed non-uniformly around the evolving ripples. This mesh transformation alters the node positions on the walls and the node to node distances would become non-uniform, causing the mesh quality to deteriorate.

A node shuffle algorithm \citep{Hewett2017} was employed to preserve the mesh quality by uniformly distributing the nodes along the perimeter of the moving silica boundary. A similar code was used for implementing this algorithm in the current simulations, with a shuffling factor of $\xi = 0.025$ (a parameter which restricts the distance that a node can be shuffled between adjacent nodes). However, instead of anchoring a node at the stagnation point (as in the eroding cylinder geometry \citep{Hewett2017}), the two end nodes of the lower boundary were fixed in the horizontal direction (constant $x^+$).

The node shuffle was applied after calculating the new node positions (from Equation~\ref{eqn:EquationF}), with a user-defined function, and before deforming the mesh with the linearly elastic solid model (Equations~\ref{eqn:EquationGa} to \ref{eqn:EquationGc}).

\subsection{Mesh quality}
The quality of the mesh was important to first obtain from its initial generation and then retain throughout the simulation where deformations due to the moving boundaries affected the mesh quality. A key metric for assessing this mesh quality is the skewness of the finite volume cells, which can be expressed as 
\begin{equation}\label{eqn:EquationE}
\text{Cell skewness} = \max \frac{\lvert \theta_\text{interior} - \frac{\pi}{2} \rvert}{ \frac{\pi}{2}}
\end{equation}
where $\theta_\text{interior}$ is the interior angle of each corner of a cell.

The skewness of cells should normally not exceed $\num{0.85}$ to avoid erroneous results, values of $\num{0.5} - \num{0.8}$ are generally acceptable and a mesh with a maximum cell skewness in the range of $\num{0.8} - \num{0.95}$ is typically regarded as poor.

Cell skewness of the initial generated mesh is zero as all cells are rectangular cuboid (Figure~\ref{fig:Figure2a}). Skewness in cells increase as they twist and deform in accordance to the evolving fluid-silica interface boundary.

Another consideration for the mesh quality is the local variations in cell volumes between finite volumes. Cell volumes gradually reduce closer to the boundary where the momentum and concentration boundary layers are resolved.  This gradual reduction in cell volumes near the boundary is mostly preserved throughout the mesh deformations as shown in Figure~\ref{fig:Figure2}.

\section{Results}
First, we validate the approach of modelling colloidal particle deposition using an Eulerian model by comparing results with an existing experiment of polystyrene particle deposition in Poiseuille flow \citep{Yang1999}. The two primary cases were investigated with the evolution of a pipe surface contracting due to silica particle deposition: (1) with an initial sinusoidal-shaped bump; and (2) with a series of five bumps forming a quintuplet.

\subsection{Validation of Eulerian deposition model}\label{sec:SectionE}
The deposition model used in the simulations was first validated against an analytical solution of particle deposition between parallel plates. We compared our results against experimental data \citep{Yang1999} where a dilute concentration of monodispersed polystyrene particles were immersed within water and deposited onto a plate in a parallel plate flow cell. The inlet velocity was $u_\text{inlet} = \SI{2.19}{\milli\meter\per\second}$ and the water properties were evaluated at $T = \SI{20}{\celsius}$ yielding a $\text{Re} = 1.31$. The particles had a diameter of $d_p = \SI{783}{\nano\meter}$ with a density of $\rho_p = \SI{1050}{\kilogram\per\meter\cubed}$.

The Smoluchowski-Levich approximation uses the analytical solution to the convection-diffusion equation under steady flow. This approximation assumes that the hydrodynamic and van der Waals interactions between the particles and wall are counterbalanced, that particles follow fluid streamlines and ignores particle deposition due to interception \citep{Elimelech1995}. The corresponding particle flux on the plate is
\begin{equation}\label{eqn:EquationB}
j_\text{SL}(x) = \dfrac{D_B c_\text{bulk}}{\Gamma(\frac{4}{3}) r_p} \left( \dfrac{2 b \text{Pe}}{9 x} \right) ^{1/3}
\end{equation}
where $r_p$ is the particle radius, $x$ the distance from the entrance in the streamwise direction and $\text{Pe}$ the Peclet number. 

The Peclet number for parallel plate flow is defined as
\begin{equation}
\text{Pe} = \frac{3 u_\text{mean} {r_p}^3}{2 b^2 D_B}
\end{equation}
where $u_\text{mean} = u_\text{inlet}$ is the mean velocity of the flow between the plates.

The corresponding dimensionless deposition velocity for the Smoluchowski-Levich approximation is
\begin{equation}
u_\text{SL}^+ = \dfrac{j_\text{SL}}{c_\text{bulk} u^*}
\end{equation}

\begin{figure}[t] \centering
  \includegraphics{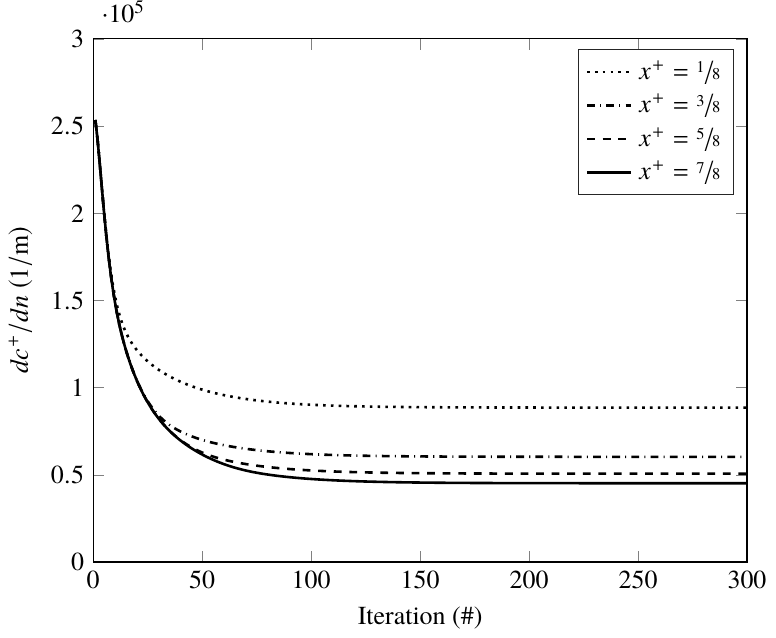}
  \caption{Iterative convergence of concentration gradient at several points equidistant along the plate, in the streamwise direction, for Mesh $\num{232} \times \num{50}$.\label{fig:Figure3}}
\end{figure}

The critical parameter to accurately capture was the concentration gradient and this parameter was converged for each run. Figure~\ref{fig:Figure3} shows that the concentration gradient converged after $\approx\num{125}$ iterations for several points along the plate.

A gradual decline of wall normal concentration gradient, and consequently particle deposition, along the length of the plate was found and is shown in Figure~\ref{fig:Figure3} (demonstrated by the non-uniform spacings between the converged concentration gradients of the equally spaced points along the plate). This non-linear relationship of $j(x^+)$ follows a $1/3$ power law as predicted from the Smoluchwoski-Levich approximation in Equation~\ref{eqn:EquationB}. The reduction in particle flux along the plate is a result of the perfect sink model applied to the wall boundaries such that the bulk concentration gradually reduces from $c_\text{bulk}$ (inlet concentration) in the streamwise direction.

Mesh convergence was obtained by refining the computational mesh and a summary of the results are shown in Table~\ref{tab:TableA}. The number of cells in the streamwise direction had a small impact on the flow field solution as the flow was unidirectional and fully developed within the primary region where deposition occurred (after the developing region near the entrance and before the exit effects). Increasing the number of cells in the wall normal direction had a greater influence on the accuracy of the concentration field as the flow had a non-zero gradient normal to the wall. Particle flux $j$ on the walls was calculated from the local concentration gradient (Equation~\ref{eqn:EquationA}) and therefore resolving the concentration boundary layer was crucial for obtaining accurate results. The simulations were within \SI{0.3}{\percent} of the analytical solution for the two most refined meshes as shown in Table~\ref{tab:TableA}.

\begin{figure}[t] \centering
  \includegraphics{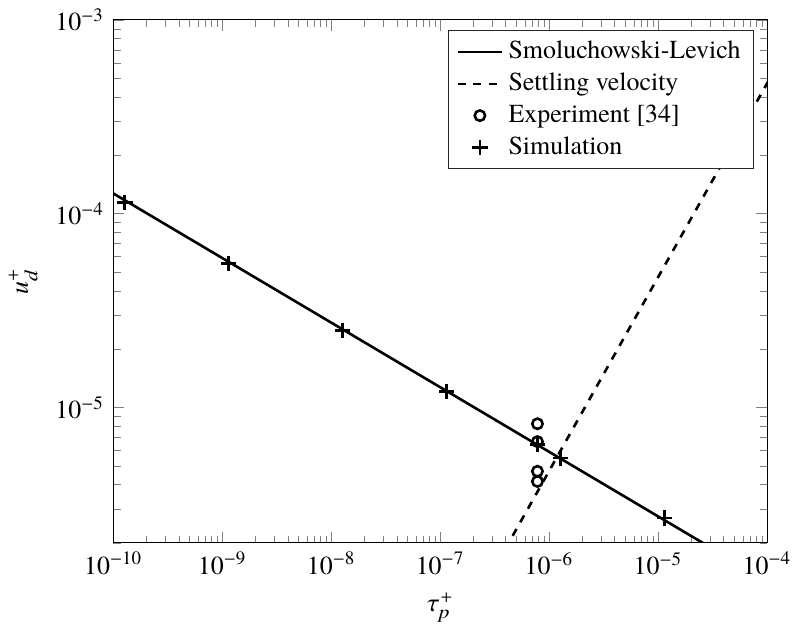}
  \caption{Dimensionless polystyrene particle deposition velocities between smooth parallel plates (at $x/L = 0.5$ with $L = \SI{0.076}{\metre}$), without obstructions, for a range of diameters (increases from left to right).\label{fig:Figure4}}
\end{figure}

A parametric study of the dimensionless particle relaxation time $\tau_p^+$ was performed with a range of particle diameters from $d_p = \SI{10}{\nano\meter}$ to $\SI{3}{\micro\meter}$. The simulation results closely matched the analytical solution of the Smoluchowski-Levich approximation across the entire range of $\tau_p^+$ as shown in Figure~\ref{fig:Figure4}.
Particle deposition $u_d^+$ was greater for smaller particles (small $\tau_p^+$) than larger particles (large $\tau_p^+$) because Brownian diffusion is inversely proportional to particle size $d_p$ (Equation~\ref{eqn:EquationC}).

Gravitational effects are significant for relatively large particles and a settling velocity can be evaluated with
\begin{equation}\label{eqn:EquationH}
u_s = \dfrac{(\rho_p - \rho) g {d_p}^2}{\num{18} \mu}
\end{equation}
where $g$ is the acceleration due to gravity which acted downwards onto the plate. The settling velocity is also non-dimensionalised for comparison with $u_d^+$ using the shear velocity such that $u_s^+ = u_s/u^*$. The total rate of deposition is a combination of all the transport mechanisms and $u_s^+$ can either enhance or reduce $u_d^+$ but the deposition velocity is bounded by $u_d^+ \geqslant 0$. The contribution of $u_s^+$ is positive when $\rho_p > \rho$ (enhancing deposition; gravity dominated) and negative when $\rho_p < \rho$ (reducing deposition; buoyancy driven).

The $u_s^+$ is negligible for small particles (small particle mass $m_p$) and significant for large particles (large $m_p$), following a square power law on $d_p$ (Equation~\ref{eqn:EquationH}). Both the Smoluchowski-Levich approximation and our simulations ignored gravitational effects, and we assume that these effects are negligible for the $\tau_p^+$ range of interest.

A range of $u_d^+$ was measured in the experiments \citep{Yang1999} as shown in Figure~\ref{fig:Figure4}. The interaction between the particles and wall was adjusted between their runs by changing the pH and ionic strength of solution, altering the zeta potentials of the surfaces and subsequently the deposition efficiency. Deposition under favourable conditions gave the highest $u_d^+$ and deposition was greatly reduced for unfavourable particle-wall interaction conditions. These near-wall interactions were not included in our simulations.

The Smoluchowski-Levich approximation and our simulation underestimated the maximum deposition rate from the experiment \citep{Yang1999} by $\SI{22}{\percent}$ which was partway between $u_\text{SL}^+$ and $u_\text{SL}^+ + u_s^+$, indicating that gravity played a role in the deposition process of these polystyrene particles. However, the total $u_d^+$ is not a linear superposition of the components $u_\text{SL}^+$ and $u_s^+$. Instead, the inertia of particles and their velocity history is important for accurately modelling both gravity and Brownian diffusion simultaneously.

Our primary simulations are exclusively in the diffusion dominated transport regime where it is well established \citep{Guha2008} that modelling only the Brownian diffusion is sufficient for accurate results.

\subsection{Laminar subdomain from turbulent pipe flow}
As mentioned in Section~\ref{sec:SectionA} of the methods, a laminar subdomain was created for modelling the pipe flow on the spatial scale of the silica ripples. The Reynolds number, based on the pipe diameter, was $\text{Re} = \num{28000}$ using a dynamic viscosity of $\mu = \SI{1.003e-3}{\kilogram\per\metre\per\second}$ and density $\rho = \SI{998.2}{\kilogram\per\metre\cubed}$; yielding a kinematic viscosity of $\nu = \SI{1.005e-6}{\metre\squared\per\second}$.

The bulk of the pipe flow was in the turbulence flow regime; however, our deforming mesh simulations were restricted to the laminar viscous sublayer. A preliminary simulation including turbulence with the $k-\epsilon$ turbulence model was undertaken to establish the fluid velocity and particle concentration profiles across the complete pipe geometry. The wall shear stress on this smooth pipe surface, without any protrusions, was $\tau_{w,\text{smooth}} = \SI{10.5}{\pascal}$; yielding a friction velocity of $u^* = \sqrt{\tau_{w,\text{smooth}}/\rho} = \SI{0.102}{\metre\per\second}$.

The colloidal silica particles had a monodispersed diameter of $d_p = \SI{21}{\nano\metre}$; yielding a relatively low Brownian diffusion value of $D_B = \SI{2.05e-11}{\metre\squared\per\second}$. The silica particles had a density of $\rho_p = \SI{1500}{\kilogram\per\metre\cubed}$; yielding a dimensionless particle relaxation factor of $\tau_p^+ = \num{3.83e-7}$.

\begin{figure}[t] \centering
  \includegraphics{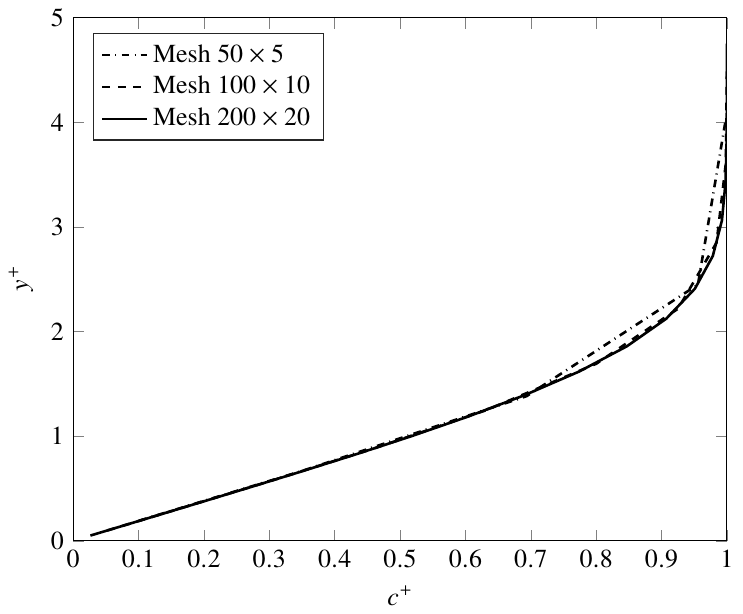}
  \caption{Dimensionless particle concentration $c^+$ within the full-length laminar subdomain for the pipe flow (at $x/L = 0.5$ with $L = \SI{1}{\metre}$) with several mesh resolutions. The bulk of the pipe flow was turbulent with a Reynolds number of $\text{Re} = \num{28000}$, and particles had a diameter of $d_p = \SI{21}{\nano\metre}$.\label{fig:Figure5}}
\end{figure}

The concentration boundary layer at the centre of the full-length subdomain ($L = \SI{1}{\metre}$) is shown in Figure~\ref{fig:Figure5} for several mesh resolutions to demonstrate mesh convergence.
This concentration profile was then applied to the short subdomain ($L_\text{sub}^+ = 50$) as a boundary condition for the evolving mesh simulations.

The silica deposition velocity (Equation~\ref{eqn:EquationD}) of the smooth surface was $u_{d,\text{smooth}} = \SI{1.1e-6}{\metre\per\second}$, whereas the settling velocity (Equation~\ref{eqn:EquationH}) was $u_s = \SI{1.2e-10}{\metre\per\second}$. The small contribution of deposition from settling compared to Brownian diffusion, with $u_s / u_{d,\text{smooth}} = \mathcal{O}(10^{-4})$, along with the small particle relaxation $\tau_p$, supports the assumption of neglecting inertial effects of the particles. Furthermore, speed of the silica deposition compared to the bulk flow was negligible ($u_{d,\text{smooth}} / u_\text{bulk} = \mathcal{O}(10^{-6})$) such that the exchange of mass from the fluid to the surface had a negligible impact on the mass fraction of silica in the bulk flow.

\subsection{Single ripple}
Simulations of the single ripple case were undertaken until the mesh quality degraded from its initially high quality state; ten mesh deformation time steps were simulated with $\Delta t^+ = \num{2500}$ yielding a final time of $t_f^+ = \num{25000}$. The solver continued to model the evolving boundary for many time steps after those reported in this paper, however the increasing skewness of cells degraded the accuracy of the solution.

\begin{figure}[t] \centering
  \begin{subfigure}{\linewidth}\begin{flushright}
    \includegraphics{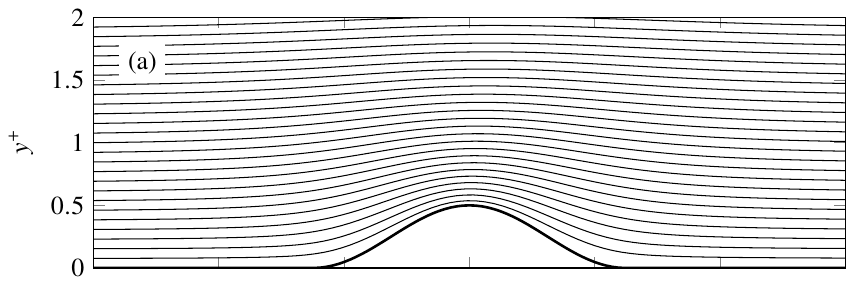}\hspace*{1.85pt}%
    \phantomcaption{\label{fig:Figure6a}}
  \end{flushright}\end{subfigure}\hfill%
  \begin{subfigure}{\linewidth}\begin{flushright}
    \includegraphics{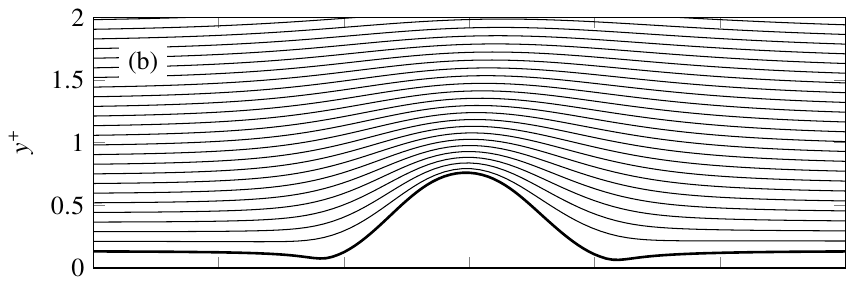}\hspace*{1.85pt}%
    \phantomcaption{\label{fig:Figure6b}}
  \end{flushright}\end{subfigure}\hfill%
  \begin{subfigure}{\linewidth}\begin{flushright}
    \includegraphics{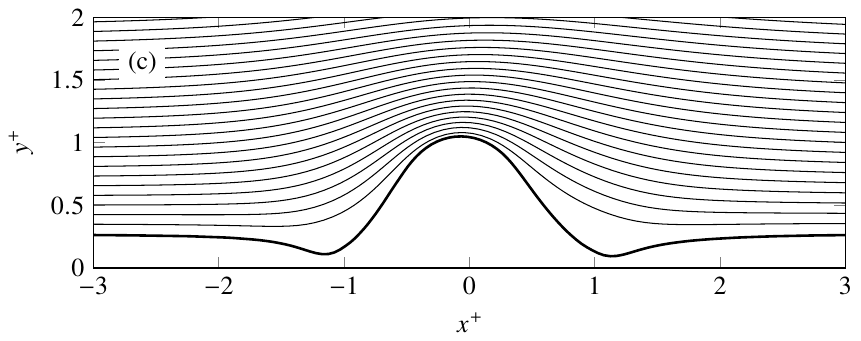}%
    \phantomcaption{\label{fig:Figure6c}}
  \end{flushright}\end{subfigure}
  \caption{Pathlines around a growing silica ripple at: (\subref{fig:Figure6a}) $t^+ = 0$; (\subref{fig:Figure6b}); $t^+ = \num{12500}$ and (\subref{fig:Figure6c}) $t^+ = \num{25000}$ with $\Delta t^+ = \num{2500}$. Flow is from left to right. The bottom solid line outlines the lower boundary of the computational domain: the deposited silica layer. Starting positions of the pathlines are equally spaced in the vertical direction and are offset by the far field bed height.\label{fig:Figure6}}
\end{figure}

The developing flow field is shown with pathlines in Figure~\ref{fig:Figure6}; illustrating an approximately symmetrical flow field about the ripple centre at $x^+ = 0$.
Fluid remained attached to the lower boundary wall without shear separation. The silica bed height upstream of the ripple matched the height of the silica downstream of the ripple, showing that the influence of the ripple on particle deposition was localised at the protrusion.

The highest fluid velocity was found at the crest of the ripple (closely spaced pathlines) and the lowest velocity in the troughs of the ripple (sparsely spaced pathlines). This trend is similar to the local growth rate of the silica bed height: highest growth at the crest and lowest growth in the troughs.

The assumption of ignoring the reduction in particle concentration along the plate due to particle deposition is supported by the fact that the particles were primarily transported out of the domain; and only a small fraction, $\mathcal{O}(10^{-8})$, of the silica deposited onto the pipe wall.

\begin{figure*}[t] \centering
  \includegraphics{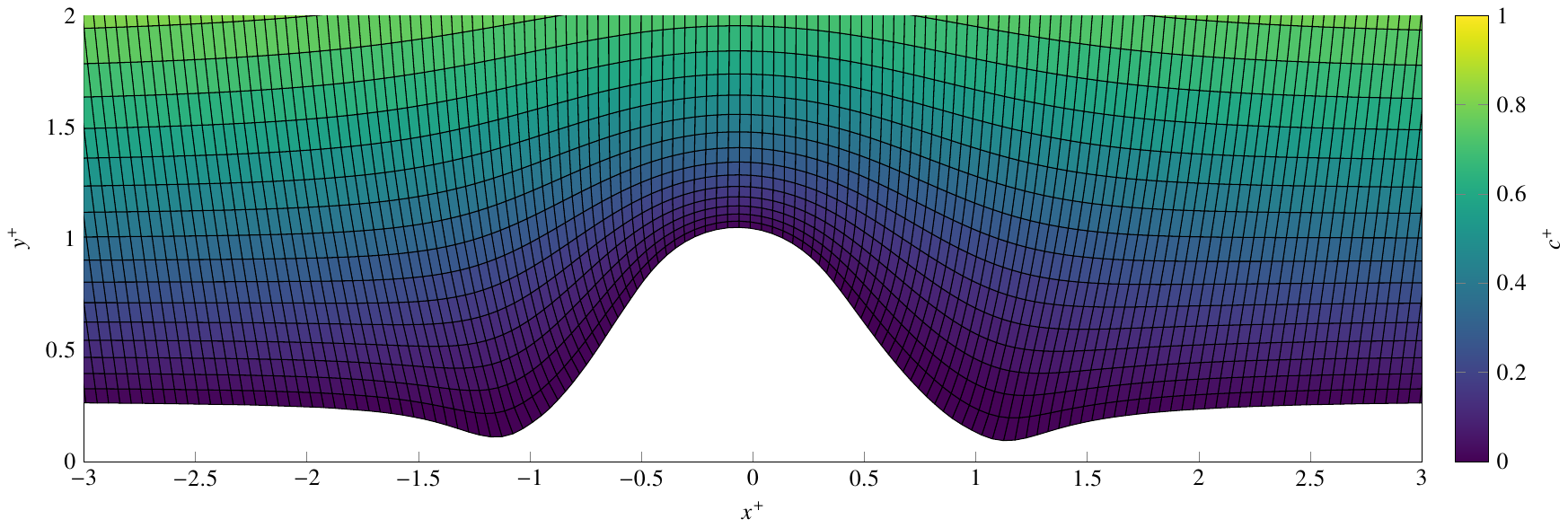}
  \caption{Close-up of dimensionless silica particle concentration $c^+$ around an evolving ripple at $t^+ = \num{25000}$ with $\Delta t^+ = \num{2500}$ (after ten mesh deformation steps). The computational mesh is shown with solid lines.\label{fig:Figure7}}
\end{figure*}

Concentration of silica $c^+$ within the flow field at the end of the simulation is shown in Figure~\ref{fig:Figure7}.
The nodes are equally spaced along the lower boundary as enforced from the node shuffle algorithm (Section~\ref{sec:SectionD}). Interior nodes were displaced with the linearly elastic solid model (Section~\ref{sec:SectionC}) which led to a low cell skewness as the mesh absorbed the deformation of the boundary.

Heights of the cells near the moving fluid-silica interface remain relatively uniform across the entire length of the domain including along the ripple (Figure~\ref{fig:Figure7}). Node spacings normal to this moving boundary were stretched at the troughs ($x^+ \approx \pm 1$) and compressed at the crest ($x^+ \approx 0$); which indirectly correspond to larger and smaller boundary layers respectively. Consequently, the resolution of the concentration boundary layer was retained along the moving boundary providing a consistent degree of accuracy for the solution. Furthermore, this boundary layer resolution was preserved throughout the mesh deformation steps over time as shown in Figure~\ref{fig:Figure2}.

Mesh convergence was quantified by comparing profile evolutions among a number of grid resolutions by varying the number of streamwise and wall normal cells. Convergence of the mesh was achieved for the $1000 \times 40$ grid and this mesh was used in both the single ripple and multiple ripple (surface roughness model) cases.

\begin{figure}[t] \centering
  \includegraphics{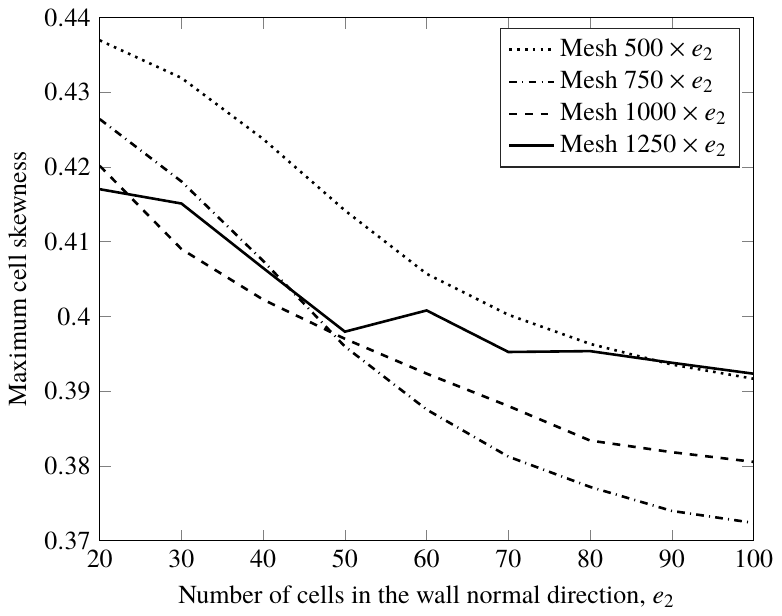}
  \caption{Mesh quality measured with maximum cell skewness for a range of grid resolutions ($\text{streamwise} \times \text{wall normal}$) at $t^+ = \num{5} \Delta t^+$ where $\Delta t^+ = \num{5200}$ (after five mesh deformation steps).\label{fig:Figure8}}
\end{figure}

Cell skewness of the mesh was evaluated with Equation~\ref{eqn:EquationE} and the maximum skewness for the mesh resolutions investigated are shown in Figure~\ref{fig:Figure8}.
The mesh quality remains at a good level after several mesh deformation steps with the maximum cell skewness below $\num{0.44}$ for all grids. There appears to be no trend of skewness with the number of streamwise cells $e_1$, but the mesh quality improves with increasing the number of wall normal cells $e_2$.

All of the simulations with a deforming mesh utilised the node shuffle algorithm outlined in Section~\ref{sec:SectionD}. This algorithm evenly distributed the nodes on the boundary and had a significant positive influence on the mesh quality. For comparison, without the node shuffle a maximum cell skewness of $\num{0.77}$ was found for the $\num{1000} \times \num{40}$ grid, compared with $\num{0.40}$ while using the algorithm (Figure~\ref{fig:Figure8}).

\begin{figure}[t] \centering
  \begin{subfigure}{\linewidth}\begin{flushright}
    \includegraphics{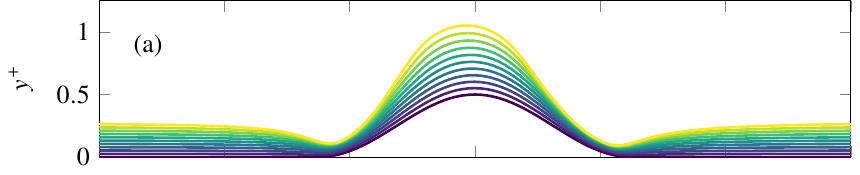}\hspace*{4.8pt}%
    \phantomcaption{\label{fig:Figure9a}}
  \end{flushright}\end{subfigure}\hfill%
  \begin{subfigure}{\linewidth}\begin{flushright}
    \includegraphics{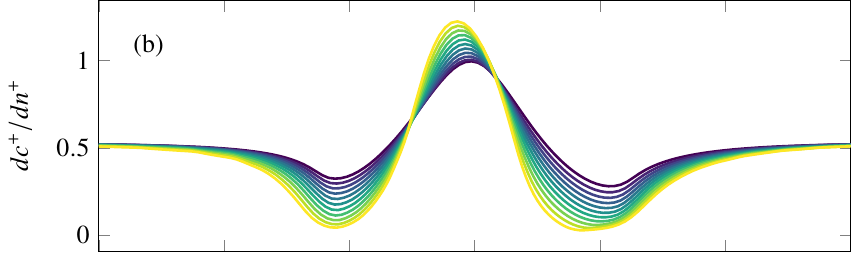}\hspace*{4.8pt}%
    \phantomcaption{\label{fig:Figure9b}}
  \end{flushright}\end{subfigure}\hfill%
  \begin{subfigure}{\linewidth}\begin{flushright}
    \includegraphics{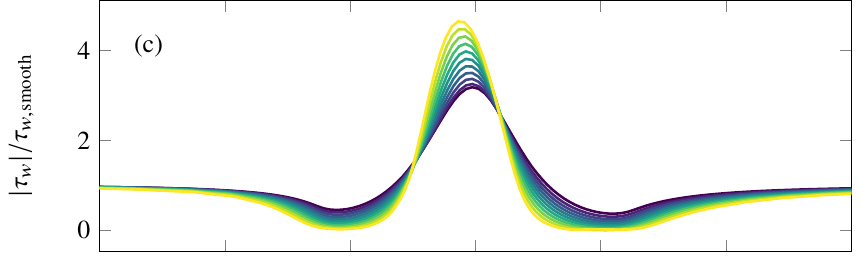}\hspace*{4.8pt}%
    \phantomcaption{\label{fig:Figure9c}}
  \end{flushright}\end{subfigure}\hfill%
  \begin{subfigure}{\linewidth}\begin{flushright}
    \includegraphics{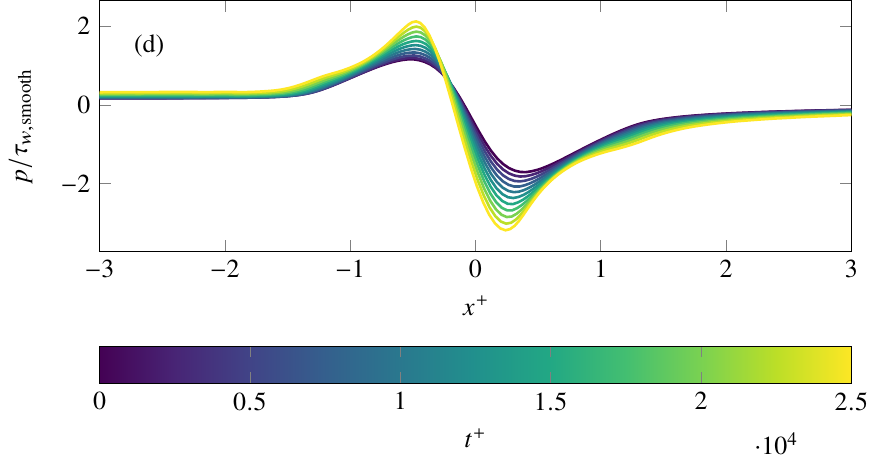}%
    \phantomcaption{\label{fig:Figure9d}}
  \end{flushright}\end{subfigure}
  \caption{Evolution of colloidal silica deposition on an initial bump within the laminar viscous sublayer of turbulent pipe flow at $\text{Re} = \num{28000}$. Time-dependent variables shown at the fluid-silica interface are: (\subref{fig:Figure9a}) height of silica deposition, (\subref{fig:Figure9b}) dimensionless particle concentration gradient, (\subref{fig:Figure9c}) wall shear stress and (\subref{fig:Figure9d}) relative pressure.\label{fig:Figure9}}
\end{figure}

The time evolution of the rising silica bed height is shown in Figure~\ref{fig:Figure9a}; from the initially prescribed sinusoidal profile to the final raised boundary profile.
In addition to the matching bed height upstream and downstream of the ripple, as commented above, the deposition rate far from the ripple remains constant over time.

Particle deposition rate, and consequently the fluid-silica interface velocity, is directly related to the concentration gradient and Figure~\ref{fig:Figure9b} corresponds to this deposition rate. This gradient $dc^+/dn^+$ highlights where the bed height was currently rising over time. The deposition rate at the troughs of the ripple significantly decreased over time towards zero whereas the deposition rate at the crest increased.

Consider a rate of silica deposition around the ripple, $\dot{m}$, as a function of the area under $dc^+/dn^+$ such that
\begin{equation}
\dot{m} = \int_{-3}^{3} D_B \frac{dc^+}{dn^+} dx^+
\end{equation}
which was calculated with numerical integration (by evaluating the curves in Figure~\ref{fig:Figure9b} with cubic splines). The integration limits on $x^+$ are arbitrary, and were chosen as $x^+ = -3$ and $x^+ = 3$ because $dc^+/dn^+$ was relatively constant outside of this range. This value $\dot{m}$ can be normalised with the rate of silica deposited in the absence of the bump
\begin{equation}
\dot{m}_\text{smooth} = \int_{-3}^{3} D_B \left. \frac{dc^+}{dn^+} \right|_\text{far field} dx^+
\end{equation}
where $\left. dc^+/dn^+ \right|_\text{far field} = 0.532$ was taken as the far field concentration gradient found both upstream and downstream of the bump. The normalised rate of silica deposition $\dot{m}/\dot{m}_\text{smooth}$ provides a quantitative comparison of the silica deposition over time as the ripple transforms its shape.

Initially $\dot{m}/\dot{m}_\text{smooth} = 0.98$ which is near unity indicating that the prescribed sinusoidal bump had a negligible effect on the overall deposition rate. However, $\dot{m}/\dot{m}_\text{smooth}$ decreases over time with an approximately linear relationship and the final ratio at the end of the simulation was $\dot{m}/\dot{m}_\text{smooth} = 0.78$; showing that fewer particles deposited along the deformed ripple when compared to both the initial bump and the smooth surface.

The magnitude of wall shear stress $|\tau_w|$ had its global maximum value at the crest of the bump for all time steps as shown in Figure~\ref{fig:Figure9c}, and had local minima in the troughs: consistent results with the pathline spacings (Figure~\ref{fig:Figure6}). The $|\tau_w|$ is near zero in the troughs of the ripple towards the end of the simulation, suggesting that the laminar shear layer was on the verge of detaching; and subsequently the flow field was near recirculation in these regions. The relative pressure along the boundary is shown in Figure~\ref{fig:Figure9d} where a positive $p$ was found on the upstream face and a negative $p$ was found on the downstream face of the ripple for all time steps.

\subsection{Surface roughness model}
Microscopic surface roughness was analysed by simulating a series of sinusoidal protrusions in the Couette flow domain. The same mesh and number of time steps with $\Delta t^+ = \num{2500}$ ($t_f^+ = \num{25000}$) were undertaken to allow a direct comparison with the single ripple case.

\begin{figure*}[t] \centering
  \begin{subfigure}{\linewidth}\begin{flushright}
    \includegraphics{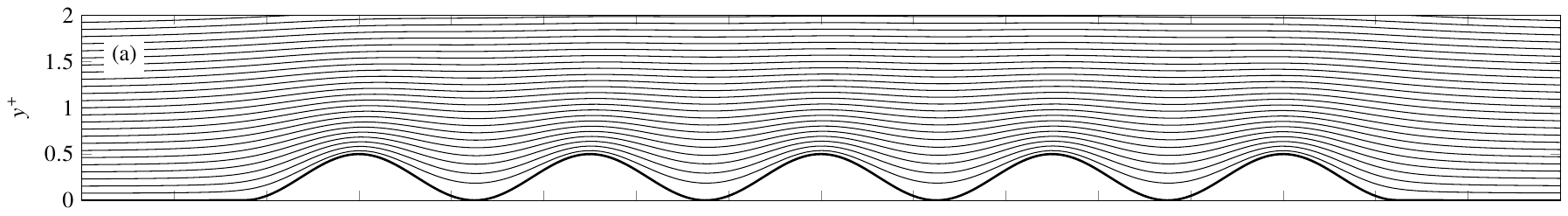}\hspace*{1.85pt}%
    \phantomcaption{\label{fig:Figure10a}}
  \end{flushright}\end{subfigure}\hfill%
  \begin{subfigure}{\linewidth}\begin{flushright}
    \includegraphics{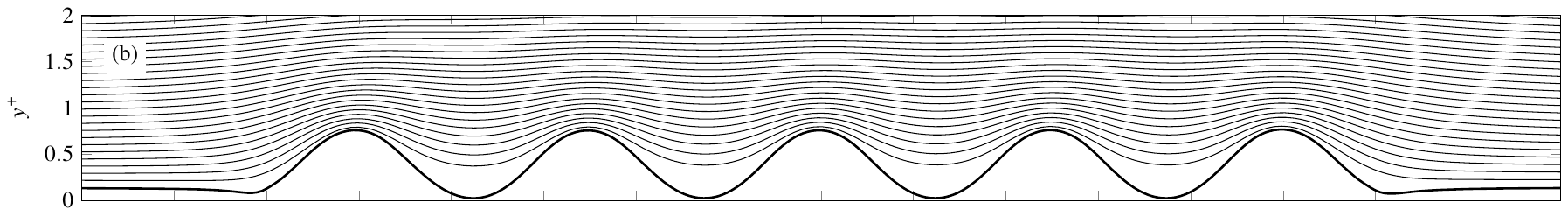}\hspace*{1.85pt}%
    \phantomcaption{\label{fig:Figure10b}}
  \end{flushright}\end{subfigure}\hfill%
  \begin{subfigure}{\linewidth}\begin{flushright}
    \includegraphics{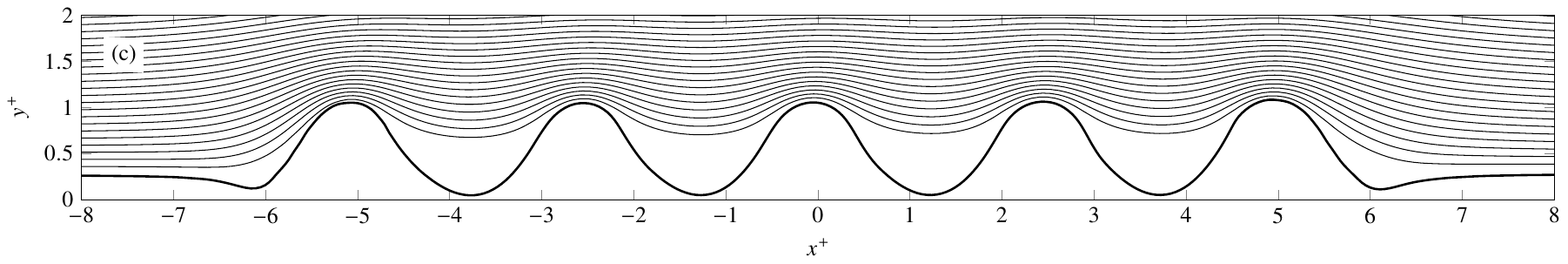}%
    \phantomcaption{\label{fig:Figure10c}}
  \end{flushright}\end{subfigure}
  \caption{Pathlines around an evolving quintuplet of silica ripples at: (\subref{fig:Figure10a}) $t^+ = 0$; (\subref{fig:Figure10b}) $t^+ = \num{12500}$; and (\subref{fig:Figure10c}) $t^+ = \num{25000}$ with $\Delta t^+ = \num{2500}$. The bottom solid line outlines the fluid-silica interface. Starting positions of the pathlines are equidistant above the silica bed.\label{fig:Figure10}}
\end{figure*}

The flow field surrounding the block of ripples (upstream of the first and downstream of the last ripple centre), as shown in Figure~\ref{fig:Figure10}, closely matches the flow for the single ripple case (Figure~\ref{fig:Figure6}).
The flow dynamics within the set of ripples had a minor effect outside of this set. Similarly, the flow in the troughs between each crest ($x^+ \approx 0, \pm 2\tfrac{1}{2}, \pm 5$) appears consistent across these four repeating unit blocks.

Tracer particle trajectories (represented with pathlines in Figure~\ref{fig:Figure10}) veer closer to crests than troughs, leading to greater deposition in the crests compared to the troughs. The pathlines deviate further from the troughs over time as the crests extend further from their origins; almost no colloidal silica particles reach these growing valleys to deposit.
 
\begin{figure}[t] \centering
  \includegraphics{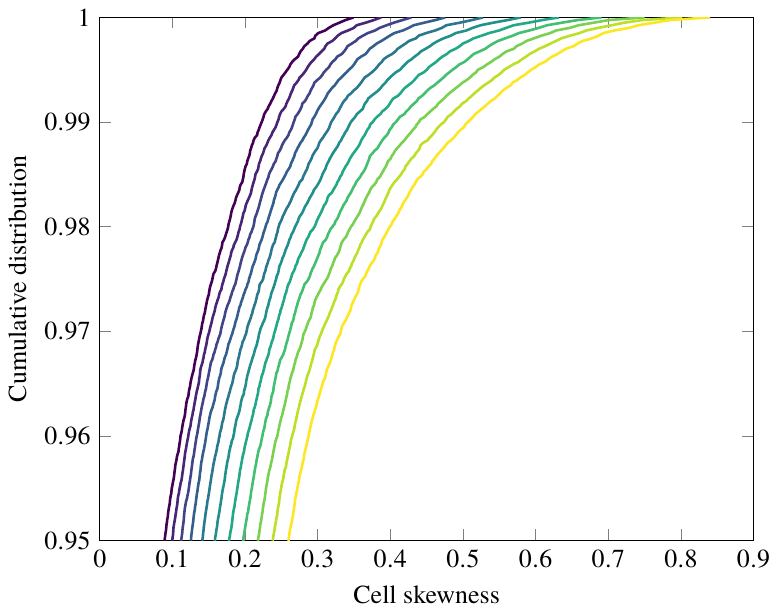}
  \caption{Empirical cumulative distribution function of cell skewness for the quintuplet of silica ripples case. The shade of the lines relate to dimensionless time and are as in Figure~\ref{fig:Figure9}.\label{fig:Figure11}}
\end{figure}

Cell skewness of the quintuplet ripple mesh progressively deteriorated as the fluid-silica interface deformed. The cell skewness of the upper fifth percentile of cells are shown in Figure~\ref{fig:Figure11} for all depositing silica time steps, and the worst affected cells were located near the crests of the outer most ripples.
The high level of maximum cell skewness experienced in the later time steps of this simulation adversely affected the accuracy that the finite volume solver could obtain. However, this data has been included in our results to illustrate the importance of mesh quality for evolving boundary problems with significant profile transformations.

The cumulative distribution of cell skewness for the single ripple case had a similar profile and followed the same trend as the quintuplet case but had a final maximum cell skewness of \num{0.40}, compared with \num{0.84}. The first layer of cells on the central ripple had similar skewness between the two cases and the skewness of cells increased sequentially for each ripple outwards in the quintuplet case. The linearly elastic mesh deformation model tended to cause cell skewness surrounding the deforming ripples (as shown in Figure~\ref{fig:Figure7}) and this trend compounded for the outer ripples in the surface roughness model.

\begin{figure*}[t] \centering
  \begin{subfigure}{\linewidth}\begin{flushright}
    \includegraphics{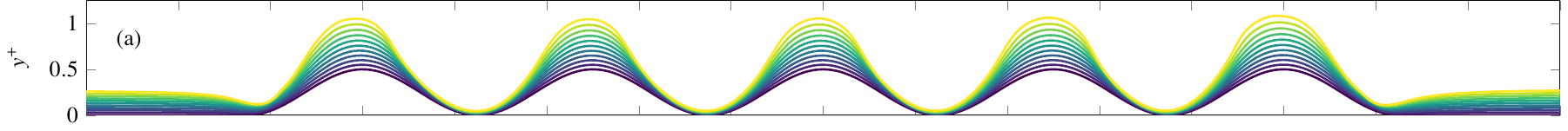}\hspace*{1.85pt}%
    \phantomcaption{\label{fig:Figure12a}}
  \end{flushright}\end{subfigure}\hfill%
  \begin{subfigure}{\linewidth}\begin{flushright}
    \includegraphics{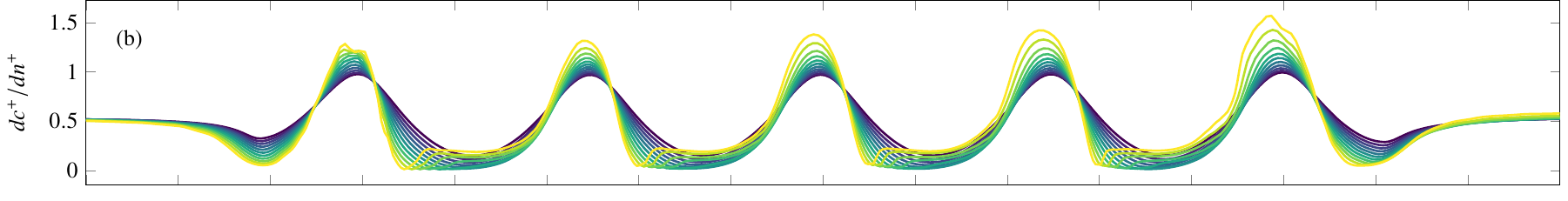}\hspace*{1.85pt}%
    \phantomcaption{\label{fig:Figure12b}}
  \end{flushright}\end{subfigure}\hfill%
  \begin{subfigure}{\linewidth}\begin{flushright}
    \includegraphics{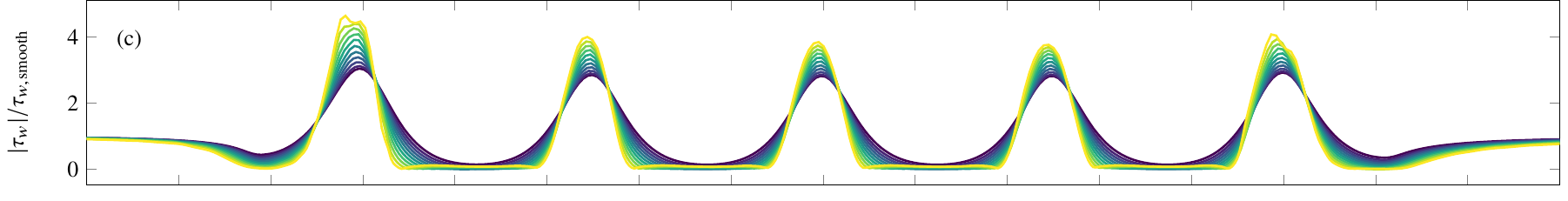}\hspace*{1.85pt}%
    \phantomcaption{\label{fig:Figure12c}}
  \end{flushright}\end{subfigure}\hfill%
  \begin{subfigure}{\linewidth}\begin{flushright}
    \includegraphics{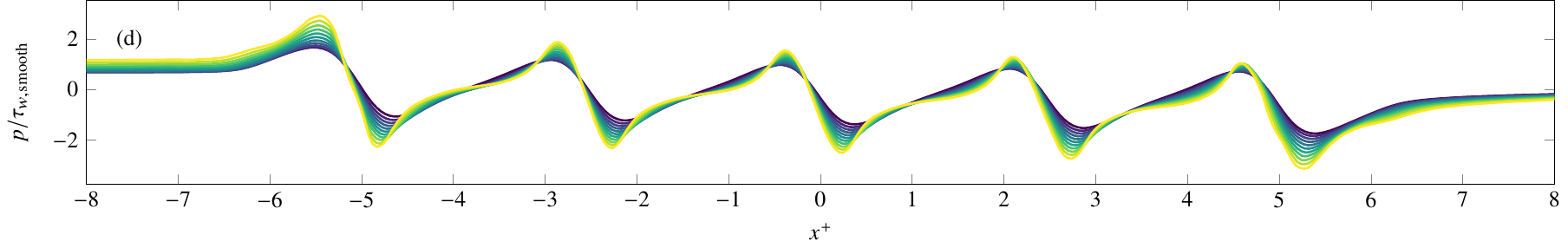}%
    \phantomcaption{\label{fig:Figure12d}}
  \end{flushright}\end{subfigure}
  \caption{Evolution of colloidal silica deposition on a quintuplet of bumps within the laminar viscous sublayer. Time-dependent variables shown at the fluid-silica interface are: (\subref{fig:Figure12a}) height of silica deposition, (\subref{fig:Figure12b}) dimensionless particle concentration gradient, (\subref{fig:Figure12c}) wall shear stress and (\subref{fig:Figure12d}) relative pressure. Shades of the lines are as in Figure~\ref{fig:Figure9}.\label{fig:Figure12}}
\end{figure*}

The evolution of the time-dependent variables for the quintuplet case, as shown in Figure~\ref{fig:Figure12}, has similarity to the single ripple (Figure~\ref{fig:Figure9}) on the upwind face of the first and on the wake side of the last ripples.
This small influence of the flow dynamics across the interior ripples on the outer faces was introduced above in describing the pathlines. Silica protrusions effectively shielded the valleys from particles within the fluid and the bed height in these regions remained relatively stagnant over time.

Irregularities in the concentration gradient profile on the outer ripples, as shown in Figure~\ref{fig:Figure12b}, in the final two to three time steps could be attributed to the high cell skewness in these regions. Typically, the accuracy of flow field variables in highly skewed cells are compromised, and the most skewed cells were found in the critical region of the first layer of cells above the fluid-silica interface. Furthermore, the accuracy of strong flow gradients is highly sensitive to cell skewness and the concentration gradient is the driving mechanism of the deforming fluid-silica interface.

The deposited silica rate for the surface roughness model also decreased with a linear relationship in time, with an initial $\dot{m}/\dot{m}_\text{smooth} = \num{0.96}$ reaching $\dot{m}/\dot{m}_\text{smooth} = \num{0.80}$ at $t^+ = 5 \Delta t^+$. The integration limits for this case were chosen as $x^+ = -7$ and $x^+ = 7$. The ratio $\dot{m}/\dot{m}_\text{smooth}$ then deviated from this trend for the second half of the simulation ($\num{12500} < t^+ \leqslant \num{25000}$) by increasing to approximately unity; this discrepancy was likely due to the increasing particle deposition in the valleys during the later time steps.

\begin{figure}[t] \centering
  \includegraphics{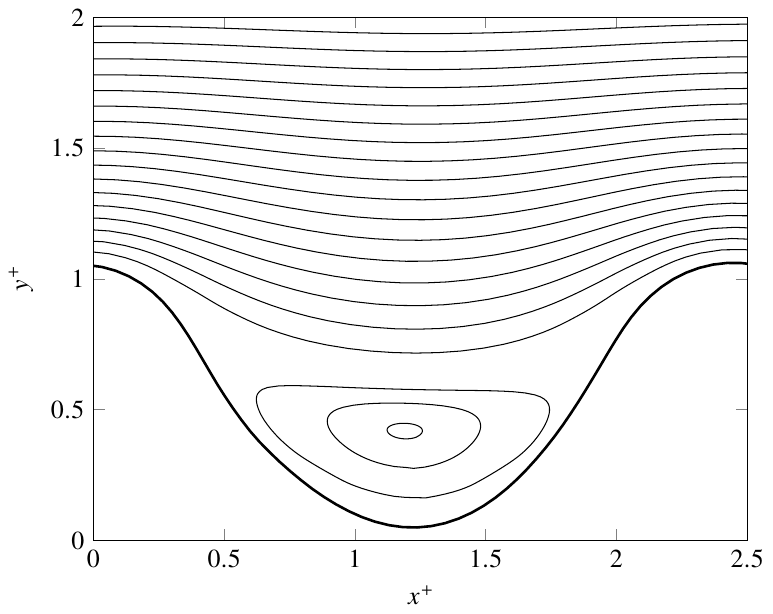}
  \caption{Close-up of pathlines for the quintuplet of silica ripples at $t^+ = \num{25000}$ (Figure~\ref{fig:Figure10}), with additional pathlines starting within the recirculation region.\label{fig:Figure13}}
\end{figure}

Flow separation occurred in the valleys between ripples for the last few time steps as shown by the wall shear stress distribution in Figure~\ref{fig:Figure12c}; yielding a recirculation region within the troughs as illustrated with pathlines in Figure~\ref{fig:Figure13}.
The shear experienced at the interface on each ripple appear similar among the interior ripples, suggesting that the flow field established the same developed profile on each crest; which follows intuition for a steady laminar flow over a repeating geometric profile. Relative pressure decreases sequentially along each ripple, with local rise and falls for each ripple, as shown in Figure~\ref{fig:Figure12d}.

\section{Discussion}

\subsection{Colloidal silica deposition}
A constant in time, and near uniform in the streamwise direction, rate of silica deposition was calculated for the smooth boundary cases in both the Poiseuille and Couette flow configurations; matching both theory \citep{Elimelech1995} and experiment \citep{Yang1999}. This relation held true because the flow was developed and in steady state (laminar flow between parallel plates, without obstructions). The same relation of silica deposition was found downstream of the protrusions after the flow redeveloped from being disturbed by the silica ripples.

The total rate of particle deposition across both the single ripple and surface roughness model geometries were lower than for the smooth surface at all time steps and reduced to $\dot{m}/\dot{m}_\text{smooth} \approx 0.8$ by the end of the simulations: the morphodynamics of the silica bed naturally reduced the deposition rate for both rough boundaries investigated. This result suggests that the optimal surface for colloidal particle deposition under steady laminar conditions is smooth rather than rough (where the rough boundary would have a greater surface area at the fluid-silica interface).

The silica ripples were non-self-stabilising: protrusions from the fluid-silica interface would grow faster than the valleys and far field smooth wall region. The opposite phenomenon was found for the boundary evolution of an eroding cylinder \citep{Hewett2017} where protrusions on the deforming boundary would erode quicker than their smooth counterparts. This non-self-stabilising effect on the surface means that if any protrusion developed, due to a numerical artefact or otherwise, it would continue to grow for the remainder of the simulation.

The quickest growth of the silica bed was found in regions of an existing protrusion and this trend correlates well with what was observed from experiments \citep{Kokhanenko2015} where mounds of silica developed. The loosely structured silica ripples, aligned normal to the direction of pipe flow (circumferential), were observed in the downstream region of the test section where the boundary layer was fully developed. Silica deposition patterns upstream, closer to the entrance, had more chaotic features and the particles were possibly transported by turbulent diffusion near the wall.

\subsection{Non-ideal particle deposition}
Dimensionless deposition velocity of the silica in the smooth pipe flow was $u_{d,\text{smooth}}^+ = \num{1.1e-5}$. Experiments under similar flow conditions and particle properties observed $u_d^+ \approx \mathcal{O}(10^{-8})$ \citep{Kokhanenko2015}: three orders of magnitude lower than what the Smoluchowski-Levich approximation predicted and what we calculated from our simulations. One hypothesis for this discrepancy between theory and experiment is that the surface roughness of the carbon steel pipes influences the deposition rate: we found that microscopic surface roughness of the form of sinusoidal ripples reduced the colloidal deposition rate by $\SI{20}{\percent}$.

Physicochemical properties of colloidal silica particles influence their deposition rate. In particular, the interaction energy between colloidal particles and a wall (combination of van der Waals and electrostatic interactions: see DLVO theory \citep{Elimelech1995}) may or may not perpetuate an energy barrier between the two surfaces \citep{Adamczyk1999}. These interactions are only significant when the particle is very close to the wall: on the order of nanometres.

Variables including the temperature, pH and concentration of dissolved minerals within the geothermal fluid affect the rate of silica precipitation and the stability of colloidal silica particles \citep{Kokhanenko2016}. The probability of a colloid attaching to a surface by forming a bond is inversely proportional to this stability. Stability of the silica colloids were measured experimentally and were found to be significantly more stable than theory \citep{Kokhanenko2016}; indicating a much lower attachment probability. Furthermore, particles may reflect or detach from the wall after impact, reducing the net amount of silica deposition.

These factors leading to a lower effective deposition rate are challenging to accurately and reliably model a priori, and therefore we focused our simulations in this paper on the hydrodynamic particle transport and boundary evolution of the rough pipe surface. Assuming a consistent deposition probability across the fluid-silica interface, enables an accurate qualitative description of the rough silica bed evolution.

\subsection{Self-similarity of a single ripple}
\begin{figure}[t] \centering
  \includegraphics{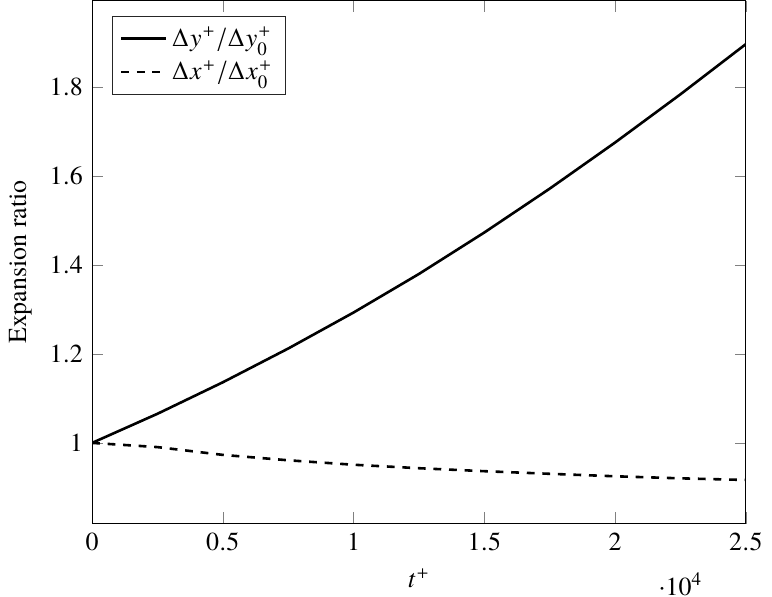}
  \caption{Expansion ratios for the single ripple case for all time steps. The height $\Delta y^+$ and width $\Delta x^+$ of the ripple are scaled by their initial values.\label{fig:Figure14}}
\end{figure}

The growth rate of the single ripple was analysed to determine if the evolution was self-similar. A dimensionless height $\Delta y^+$ was defined as the wall normal distance between the peak and the mean height of the two adjacent troughs $y^+_\text{troughs}$. Similarly, the dimensionless width $\Delta x^+$ was defined as the distance between the two troughs in the streamwise direction. The location of extrema were calculated by interpolating mesh nodes using cubic splines. These distances were then scaled by their initial values, $\Delta y_0^+ = 2 A$ and $\Delta x_0^+ = \lambda$ respectively, to define expansion ratios which are shown over time in Figure~\ref{fig:Figure14}.
As the troughs were absent in the initially prescribed sinusoidal profile, an initial width of the ripple wavelength was used instead.

The height expansion ratio $\Delta y^+ / \Delta y_0^+$ increased over time as the silica bump grew. This ratio closely followed an exponential relation in time, with
\begin{equation}\label{eqn:EquationI}
\frac{\Delta y^+}{\Delta y_0^+} = \exp{(\num{2.57e-5} t^+)}
\end{equation}
throughout the entire simulation.

In contrast, the width expansion ratio $\Delta x^+ / \Delta x_0^+$ of the silica ripple decreased for all time steps: the ripple narrowed over time. However, the rate of narrowing for the ripple was much smaller than the growth rate in the wall normal direction. This width expansion ratio approximately followed a power law relationship in time, with
\begin{equation}\label{eqn:EquationM}
\frac{\Delta x^+}{\Delta x_0^+} = 1.37 {(t^+)}^{-0.0395}
\end{equation}
after the initial first couple of mesh deformation steps.

\begin{figure}[t] \centering
  \includegraphics{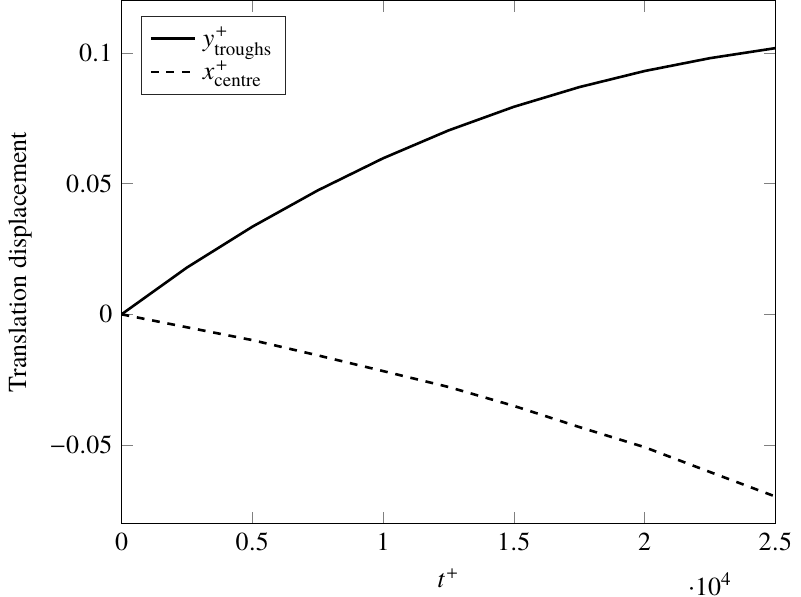}
  \caption{Translation displacements of the mean trough height and centre of bump for the single ripple case over all time steps.\label{fig:Figure15}}
\end{figure}

The evolving silica bed height also rose over time with a mean trough height of $y^+_\text{troughs}$ as shown in Figure~\ref{fig:Figure15}.
The growth rate of the troughs slowed over time as valleys formed and the colloidal silica particles veered further from the boundary (see pathlines in Figure~\ref{fig:Figure6}). This mean trough height followed a quadratic relationship in time with
\begin{equation}\label{eqn:EquationN}
y^+_\text{troughs} = -\num{1.24e-10} {(t^+)}^2 + \num{7.12e-6} t^+
\end{equation}

The crest of the ripple, shown as $x^+_\text{centre}$ in Figure~\ref{fig:Figure15}, moved upstream ($dx^+_\text{centre}/dt^+ < 0$) over time: representing a travelling wave solution. Starting at an initially motionless state, the peak (and thus ripple) translated with a constant acceleration throughout the simulation. This streamwise development of the silica bed generated a slight tilt of the ripple towards the upstream direction. This front steepening was likely caused by a non-linear dispersion relationship for the travelling wave solution. The ripple centre also followed a quadratic relationship in time with
\begin{equation}
x^+_\text{centre} = -\num{4.39e-11} {(t^+)}^2 - \num{1.67e-6} t^+
\end{equation}

The following similarity variables are defined based on the expansion ratios and translation displacements above for the single ripple case. The positions were translated by aligning the peak at the origin in the streamwise direction ($x^+_\text{centre}$) with
\begin{equation}\label{eqn:EquationJ}
\tilde{x}^+ = \frac{x^+ - x^+_\text{centre}}{1.37 {(t^+)}^{-0.0395}}
\end{equation}
and shifting the profiles down in the wall normal direction such that their mean trough heights intersect $y^+ = 0$ with
\begin{equation}\label{eqn:EquationK}
\tilde{y}^+ = \frac{y^+ - y^+_\text{troughs}}{\exp{(\num{2.57e-5} t^+)}}
\end{equation}
where $x^+_\text{centre}$ and $y^+_\text{troughs}$ are quadratically dependent on time $t^+$ with Equations~\ref{eqn:EquationM} and \ref{eqn:EquationN} respectively, and their values are shown in Figure~\ref{fig:Figure15}.

\begin{figure}[t] \centering
  \includegraphics{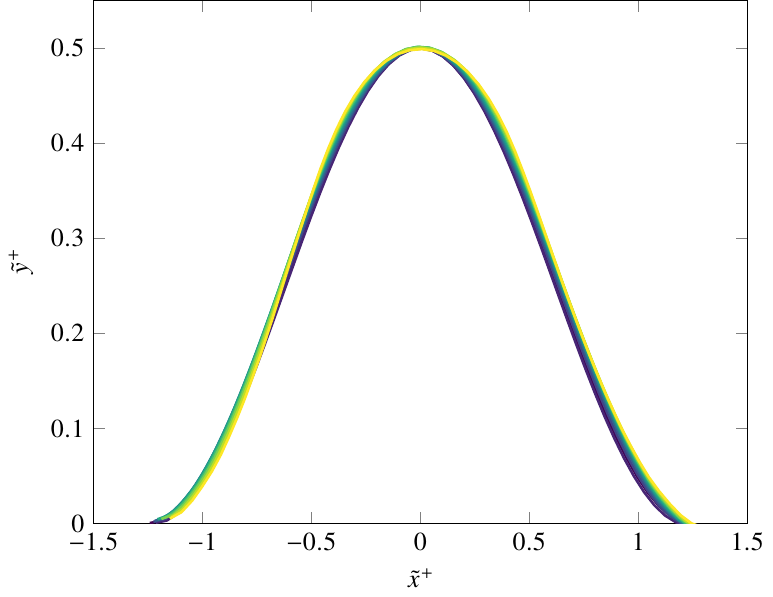}
  \caption{Transformed profiles for the single ripple case with the similarity variables $\tilde{x}^+$ (Equation~\ref{eqn:EquationJ}) and $\tilde{y}^+$ (Equation~\ref{eqn:EquationK}) applied for each time step to demonstrate the self-similar evolution. Shades of the lines are as in Figure~\ref{fig:Figure9}.\label{fig:Figure16}}
\end{figure}

The ripple profiles using these similarity variables $\tilde{x}^+$ and $\tilde{y}^+$ of each time step are shown superimposed in Figure~\ref{fig:Figure16}.
The silica bed height outside of the ripple (defined as the shape between the two local minima) is excluded from this figure.

Profiles of the silica ripple over time collapsed well onto a single master curve (Figure~\ref{fig:Figure16}); demonstrating a self-similar solution. The peak of the transformed profiles followed very closely with the master curve over time (Equation~\ref{eqn:EquationL}), matching the initial height $y^+ = 0.5$.

\subsection{Deforming mesh approach}
The ALE model is a powerful tool used for deforming mesh simulations with the distinct advantage of tracking the moving boundaries explicitly; when compared to the immersed boundary methods where the interface is captured with a model. The linearly elastic solid model (Section~\ref{sec:SectionC}) used for smoothing the interior mesh nodes is effective at maintaining mesh quality for large domain volume changes; we have previously employed this model for tracking a melting ice front \citep{Hewett2017a} where the domain size increased by an order of a magnitude. We have also used a similar smoothing mesh method, based on a diffusion model, for simulating the evolution of an eroding cylinder in cross flow \citep{Hewett2017}; where the eroding cylinder shrunk by an order of magnitude during the simulation.

\begin{figure*}[t] \centering
  \includegraphics{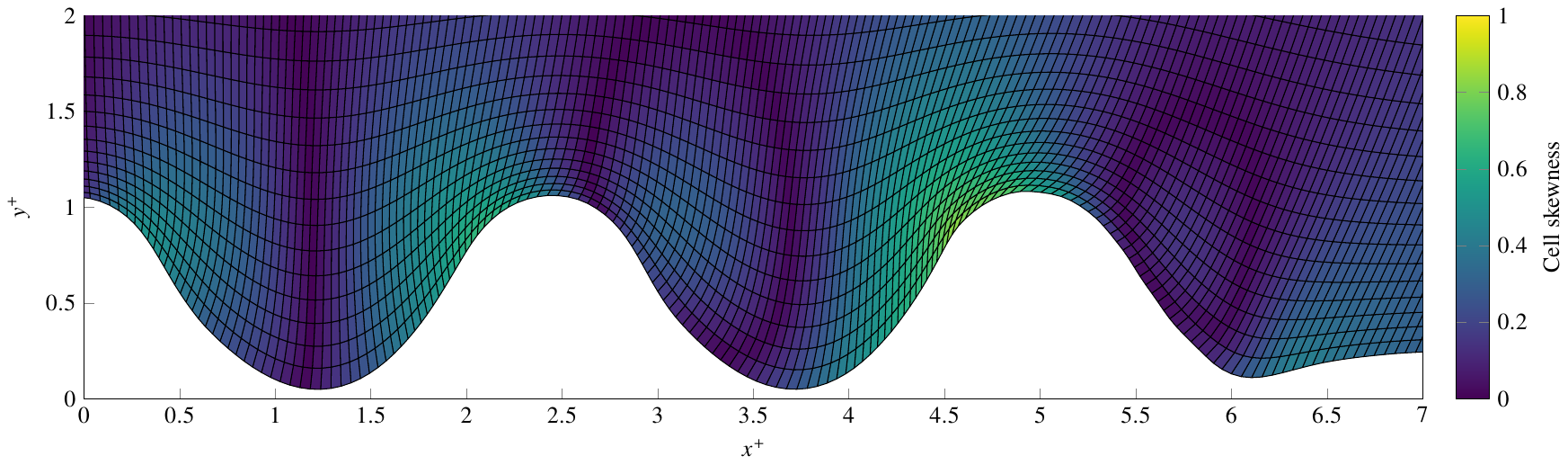}
  \caption{Close-up of cell skewness around the downstream region of the quintuplet of ripples at $t^+ = \num{25000}$ (final geometry after ten mesh deformation steps). The finite volume cells are outlined with solid lines. The upstream region ($x^+ < 0$) was the mirror image of this figure including the cell skewness values.\label{fig:Figure17}}
\end{figure*}

Cell skewness is an important metric to quantify the mesh quality. As the simulations presented in this paper were dealing with deforming meshes using evolving boundaries, the mesh quality over the duration of the simulations had to be monitored. Irregularities in the solution, particularly the wall normal particle concentration gradient $dc^+/dn^+$, were found in the final time steps of the quintuplet ripple case (Figure~\ref{fig:Figure12b}) in regions where the cell skewness was around $0.6 - 0.7$ at $x^+ \approx \pm 5$ within the first cell layer as shown in Figure~\ref{fig:Figure17}.
Similarly, irregularities were also found in the wall shear stress $\tau_w$ in Figure~\ref{fig:Figure12c} (directly proportional to the velocity gradient); whereas this issue was less prevalent in the variables without gradients (relative pressure $p$ in Figure~\ref{fig:Figure12d}), indicating that gradients were poorly calculated in highly skewed cells. The simulation was then terminated as the accuracy of the results had degraded.

The adverse effect of cell skewness on the accuracy of particle deposition was evaluated by imposing a prescribed distortion on the mesh used in the case studied earlier in Section~\ref{sec:SectionE}. The mesh was distorted with varying degrees by shifting all interior nodes in the streamwise direction with
\begin{equation}
\Delta x  = A_\text{skew} \sin \left( N_\text{skew}\frac{\pi x}{L} \right) \sin \left( \frac{\pi y}{2 b} \right)
\end{equation}
where $A_\text{skew}$ is the skew amplitude and $N_\text{skew}$ the number of half sinusoids; yielding gradual changes in cell skewness towards a peak located at the centre of the plate for odd $N_\text{skew}$.

Increasing the cell skewness at the centre of the bottom plate resulted in higher errors for the local deposition flux as shown in Table~\ref{tab:TableB}.
\begin{table}[t]
\centering\small
\caption{Effect of cell skewness on particle deposition flux at the centre of the non-evolving parallel plate with $N_\text{skew} = 1$.}
\label{tab:TableB}
\begin{tabular}{@{}S[table-format=1.1]S[table-format=1.3]S[table-format=3.1]S[table-format=2.2]@{}}
\toprule
{$A_\text{skew}$ ($\SI{}{\milli\metre}$)}     & {Cell skewness}    & {$j~(\SI{}{\#\per\centi\metre\squared~\second})$}  & {Error (\SI{}{\percent})} \\ \midrule
0.0                                           & 0.000              & 301.7                                              &                         \\
0.1                                           & 0.307              & 301.9                                              &  0.07                   \\
0.2                                           & 0.515              & 302.0                                              &  0.10                   \\
0.3                                           & 0.639              & 302.1                                              &  0.14                   \\
0.4                                           & 0.716              & 302.2                                              &  0.19                   \\
0.5                                           & 0.768              & 302.5                                              &  0.27                   \\
1.0                                           & 0.880              & 304.7                                              &  1.01                   \\
1.5                                           & 0.919              & 308.6                                              &  2.28                   \\
2.0                                           & 0.939              & 314.0                                              &  4.08                   \\
2.5                                           & 0.952              & 321.0                                              &  6.41                   \\
3.0                                           & 0.960              & 329.6                                              &  9.25                   \\
3.5                                           & 0.965              & 339.7                                              & 12.6                    \\
4.0                                           & 0.970              & 351.3                                              & 16.5                    \\ \bottomrule
\end{tabular}
\end{table}
The influence of the mesh distortion on the solution accuracy was insignificant for cell skewness values less than approximately $\num{0.7}$, whereas for a cell skewness above $\num{0.96}$ the error was over $\SI{10}{\percent}$.

Skewness in cells is a good indicator for assessing the local quality of a mesh, although the accuracy of the corresponding local solution also depends on the global mesh quality and flow dynamics. For example, with $N_\text{skew} = 9$ an error of $\SI{0.53}{\percent}$ was found for $A_\text{skew} = \SI{0.5}{\milli\metre}$ yielding the same local cell skewness of $\num{0.768}$ as in Table~\ref{tab:TableB}; demonstrating the dependence of solution accuracy on the global mesh quality. Furthermore, the dependence on flow dynamics is illustrated in Figure~\ref{fig:Figure12} for the case of the silica deposition in the surface roughness model; non-linear streamlines resulted in a greater sensitivity to the cell skewness.

The two other evolving boundary simulation papers on a melting front and an eroding cylinder mentioned above demonstrated that this deforming mesh approach functions well in deforming bodies where the moving boundary shifts in a predominately normal direction. The node shuffle algorithm (uniformly distributing nodes along the boundary) improved the mesh quality for changing curvatures (circular to rounded triangular). However, whilst the node shuffle provided a uniformly distributed boundary for the surface roughness model in this paper, the mesh twisted as the ripples evolved (as shown in Figure~\ref{fig:Figure17}). The cause of this twisting was the overall movement of nodes along the fluid-silica interface towards the centre $x^+ = 0$ as the growing ripples increased the length of this interface over time. The interior mesh was not deformed with a requirement of uniformly spaced nodes, whereas this requirement was imposed on the interface boundary with the node shuffle algorithm.

Both the linearly elastic solid model and the diffusion based model for smoothing the interior nodes of the mesh were trialled with preliminary simulations, with the former yielding a higher mesh quality as the mesh evolved. The high cell skewness near the deforming boundary was the critical limiting factor of the simulations, and improving the cell quality would enable the simulations to continue as the underlying governing models and theory remained valid for the evolved silica bed profile.

Future work could examine the use of alternative models for deforming the mesh. Remeshing, a relatively computationally expensive task, could still be avoided by smoothing the interior mesh; the general flow characteristics and requirement of resolving the concentration boundary layer remains consistent over the silica bed height evolution. A possible alternative model could include a third step which would reduce skewness within cells after node shuffling the boundary nodes and smoothing the interior cells. This step could be achieved by applying the node shuffle to the layers of cells near the moving boundary (restricted to structured grids) or by generating a new algorithm for aligning cell corners with optimal angles (applicable for both structured and unstructured grids).

\section{Conclusion}
Colloidal particle deposition of silica was simulated within the laminar viscous sublayer of turbulent pipe flow with an evolving boundary. First, the Eulerian approach of modelling colloidal particle deposition was validated against existing experiments, and the simulation results also matched closely with the Smoluchowski-Levich approximation. Then, a deforming mesh model was used for simulating the evolution of a rough silica bed surface in pipe flow and was compared with another set of experiments. We found that silica deposition was enhanced on crests of the sinusoidal protrusions with an exponential growth in time. In contrast, the silica deposition was reduced within the valleys when compared to the smooth surface case. The discrepancy between the smooth and rough wall models enlarged as the protrusions grew, ultimately reducing the overall deposition efficiency by $\SI{20}{\percent}$ over the time simulated.

The initial sinusoidal protrusions imposed on the deforming fluid-silica interface developed self-similarly over time. Skewness of the finite volume cells increased as the mesh evolved and was highly correlated with irregularities in the solution where the cell skewness reached or exceeded $0.6 - 0.7$. The silica ripples accelerated in the upstream direction as the colloidal particle deposition favoured the upstream face of each bump; this travelling wave phenomenon contributed to the skewness of cells.


\bibliography{SilicaDeposition}

\end{document}